\documentclass[10pt,aps,showpacs,floatfix,amsmath,amssymb,groupedaddress]{revtex4}
\usepackage{epsfig}
\usepackage{graphicx}
\usepackage{graphics}
\usepackage{xspace}
\usepackage{bm}
\usepackage{amssymb}
\usepackage{latexsym}
\usepackage{natbib}
\usepackage{mathrsfs}
\newcommand{\inieq}{\begin{eqnarray}}            
\newcommand{\fineq}{\end{eqnarray}}            
\newcommand{\sla}{\mskip 1.mu /\mskip-9mu}      
\newcommand{\diff}{{\rm\,d}}                    
\newcommand{\be}{\begin{equation}}
\newcommand{\ee}{\end{equation}}
\newcommand{\ba}{\begin{eqnarray}}
\newcommand{\ea}{\end{eqnarray}}

\newcommand{\g}{\gamma}
\newcommand{\nn}{\nonumber}
\def\bea{\begin{eqnarray}}
\def\eea{\end{eqnarray}}

\def\x{\mbox{\boldmath $x$}}
\def\y{\mbox{\boldmath $y$}}

\def\p{\mbox{\boldmath $p$}}

\def\q{\mbox{\boldmath $q$}}

\def\k{\mbox{\boldmath $k$}}

\def\ta{\mbox{\boldmath $\tau$}}

\begin{document}
\title{{ A relativistic model for the non-mesonic weak decay
of the ${_\Lambda^{12} C}$ hypernucleus}}
\author{Francesco Conti}
\affiliation{Dipartimento di Fisica Nucleare e Teorica,
Universit\`{a} degli Studi di Pavia and \\
Istituto Nazionale di Fisica Nucleare,
Sezione di Pavia, I-27100 Pavia, Italy}
\author{Andrea Meucci}
\affiliation{Dipartimento di Fisica Nucleare e Teorica,
Universit\`{a} degli Studi di Pavia and \\
Istituto Nazionale di Fisica Nucleare,
Sezione di Pavia, I-27100 Pavia, Italy}
\author{Carlotta Giusti}
\affiliation{Dipartimento di Fisica Nucleare e Teorica,
Universit\`{a} degli Studi di Pavia and \\
Istituto Nazionale di Fisica Nucleare,
Sezione di Pavia, I-27100 Pavia, Italy}
\author{Franco Davide Pacati }
\affiliation{Dipartimento di Fisica Nucleare e Teorica,
Universit\`{a} degli Studi di Pavia and \\
Istituto Nazionale di Fisica Nucleare,
Sezione di Pavia, I-27100 Pavia, Italy}

\date{\today}

\begin{abstract}
A  fully relativistic finite nucleus wave-function approach to the non-mesonic
weak decay of the ${_\Lambda^{12} C}$ hypernucleus is presented. The model is
based on the calculation of the amplitudes of the tree-level Feynman diagrams
for the $\Lambda N\to NN$ process and includes one-pion exchange and one-kaon
exchange diagrams. The pseudo-scalar and pseudo-vector choices for the vertex
structure are compared. Final-state interactions between each one of the
outgoing nucleons and the residual nucleus are accounted for by a complex
phenomenological optical potential. Initial $\Lambda N$ and final $NN$ short-range correlations
are included by means of phenomenological correlation functions. Numerical results are presented and discussed for
the total non-mesonic decay width $\Gamma_{nm}=\Gamma_n+\Gamma_p$, the
$\Gamma_n/\Gamma_p$ ratio, the $a_\Lambda$ intrinsic asymmetry parameter, and
the kinetic energy and angular spectra.
\end{abstract}


\maketitle


\section{Introduction}
\label{sec:intro}

The birth of hypernuclear physics dates back to 1952  \cite{HypFirst} when the
first hypernuclear fragment originated from the collision of a high-energy
cosmic proton and a nucleus of the photographic emulsion exposed to cosmic rays
was observed through its weak decays, revealing the presence of an unstable
particle: this was interpreted as due to the formation of a nucleus in which a
neutron is replaced by the $\Lambda$ hyperon, i.e., the lightest strange baryon.
A hypernucleus is a bound system of neutrons, protons, and one or more hyperons.
Only the lightest hyperon, the $\Lambda$, is stable with respect to esoenergetic
strong and electromagnetic processes in nuclear systems. Therefore, the most
stable hypernuclei are those made up of nucleons and a $\Lambda$ particle.
We denote with  $^{A}_\Lambda X$ a hypernucleus with $Z$ protons, $(A-Z-1)$
neutrons, and a $\Lambda$ ($\Lambda$-hypernucleus).

Hypernuclei represent a unique laboratory for the study of strong and weak
interactions of hyperons and nucleons through the investigation of
hypernuclear structure and decay. The $\Lambda$ particle, which does not have to
obey the Pauli principle, is an ideal low-energy probe of the nuclear
environment which allows a deepening of classical nuclear physics subjects,
such as the role of nuclear shell models and the dynamical origin of the
nuclear spin-orbit interaction. Hypernuclear physics also establishes a bridge
between nuclear and hadronic physics, since many related issues can in
principle unravel the role played by quarks and gluons partonic degrees of
freedom inside nuclei. In this direction, the study of hybrid theories
combining meson-exchange mechanisms with direct quark interactions have the
potentiality to teach us something on the confinement phenomenon, an issue
still far from being satisfactorily understood.

In $\Lambda$-hypernuclei the $\Lambda$ can decay via either a mesonic or a
non-mesonic strangeness-changing weak interaction process. In the nuclear
medium the mesonic decay, $\Lambda\to N\pi$, which is the same decay of a free
$\Lambda$, is strongly suppressed, but in the lightest hypernuclei, by the
effect of the Pauli principle on the produced nucleon, whose  momentum
($\sim$ 100 MeV/$c$) is well below the Fermi momentum. In the non-mesonic weak
decay (NMWD) the pion produced in the weak $\Lambda\to N\pi$ transition is
virtual and gets absorbed by neighbor nucleons. Then, two or three nucleons
with high momenta ($\sim$ 400 MeV/$c$) are emitted. We can distinguish between
one and two-nucleon induced decays, according to whether the $\Lambda$ interacts
with a single nucleon, either a proton, $\Lambda p \to np$ (decay width
$\Gamma_p$), or a neutron, $\Lambda n\to nn$ ($\Gamma_n$), or with a pair of
correlated nucleons, $\Lambda NN\to nNN$ ($\Gamma_2$).  Mesons heavier than the
pion can also mediate these transitions. The NMWD process is only possible in
the nuclear environment and represents the dominant decay channel in
hypernuclei beyond the $s$-shell.
The total weak decay rate is given by the sum of the mesonic ($\Gamma_m$) and
non-mesonic ($\Gamma_{nm}$) contributions:
\be \Gamma_{tot}= \Gamma_m + \Gamma_{nm}, \ee
with
\be \Gamma_{nm}= \Gamma_1+\Gamma_2, \qquad \Gamma_1=\Gamma_p+\Gamma_n. \ee

The fundamental interest in the NMWD mode is that it provides a unique tool to
study the weak strangeness changing ($|\Delta S=1|$) baryon-baryon interaction
$\Lambda N\to nN$, in particular its parity conserving part, that is much more
difficult to study with the weak  $NN \to NN$ transition, that is overwhelmed
by the parity-conserving strong $NN$ interaction. Since no stable hyperon
beams are available, the weak process $\Lambda N\to nN$ can be investigated
only with bound strange systems. The study of the inverse process
$pn \to p \Lambda$ would however be useful.

Although the relevance of the NMWD channel was recognized since the early days
of hypernuclear physics, only in recent years the field has experienced great
advances due to the conception and realization of innovative experiments and to
the development of elaborated theoretical models
\cite{stab1,rev1,rev2,rev2a,rev3,rev4,rev5}.

For many years the main open problem in the decay of hypernuclei has been the
$\Gamma_n/\Gamma_p$ puzzle, i.e., the disagreement between theoretical
predictions and experimental results of the ratio between the neutron- and
proton-induced decay widths: for all the considered hypernuclei the experimental
ratio, in the range  $\sim 0.5 \div 2$, was strongly underestimated (by about
one order of magnitude) by the theoretical results.
The $\Gamma_n/\Gamma_p$ ratio directly depends on the isospin structure of the
weak process driving the hypernuclear decay.
The analysis of the ratio is a complicated task, due to difficulties in the
experimental extractions, which require the detection of the decay products,
especially neutrons, and to the presence of additional competing
effects, such as final-state interactions (FSI) of the outgoing nucleons and
two-nucleon induced decays,  which could in part mask and modify the original
information.

In the first theoretical calculations the one-pion-exchange (OPE)
nonrelativistic picture was adopted as a natural starting point in the
description of the $\Lambda N\to nN$ process, mainly on the basis of its success
in predicting the basic features of the strong $NN$ interaction. The first OPE
models were able to reproduce the non-mesonic decay width
$\Gamma_{nm}=\Gamma_n+\Gamma_p$ but predicted too small  $\Gamma_n/\Gamma_p$
ratios \cite{L1,L3,L4,L4b,L2,L4c}. It thus seemed that the theoretical approaches
tend to underestimate $\Gamma_n$ and overestimate $\Gamma_p$. A solution of
the puzzle then requires devising dynamical effects able to increase the
$n$-induced channel and decrease the $p$-induced one.

In the following years the theoretical framework was improved including the
exchange of all the pseudo-scalar and vector mesons, in the form of a full
one-meson-exchange (OME) model, or properly simulating additional effects,
above all initial short-range correlations (SRC) and FSI, by
means of direct quark mechanisms and many-body techniques
\cite{L1,L3,L4b,L2,L8,L5,L6,L7,DI3,DI3b,DI3c,hqm,Dq1,Dq2}. In particular, the
inclusion of $K$ exchanges seems essential to improve the
agreement between theory and experiments. Only a few of these calculations have
been able to predict a sizeable increase of the $\Gamma_n/\,\Gamma_p$ ratio
\cite{L2,L4c,L8,Dq2}, but no fundamental progress has been achieved concerning
the deep dynamical origin of the puzzle.

The situation has been considerably clarified during the very last years,
thanks to considerable progress in both  experimental techniques
\cite{exp1,exp2,exp3,exp4,exp5,exp6,exp7} and theoretical
treatments \cite{L2,L4c,L8,Dq2,th1,th2,th3,th4,jun,Asth1,th5,last,Bauer:2010tk}.
From the experimental point of view, the new generation of KEK experiments has
been able to measure the fundamental observables for the $^5_\Lambda He$ and
$^{12}_\Lambda C$ hypernuclei with much more precision as compared with the
\lq\lq old'' data, also providing the first results of simultaneous
one-proton and one-neutron energy spectra, which can be directly compared
with model calculations. Very recently, it has also been possible to obtain
for the first time coincidence measurements of the nucleon pairs emitted in
the non-mesonic decay, with valuable information on the corresponding
angular and energy correlations. These new data further refine our experimental
knowledge of the hypernuclear decay rates, also allowing a cleaner and more
reliable extraction of the $\Gamma_n/\,\Gamma_p$ ratio. From the theoretical
point of view, crucial steps towards the solution of the puzzle have been
carried out, mainly through a non-trivial reanalysis of the pure experimental
results by means of a proper consideration of FSI and rescattering mechanisms,
inside the nuclear medium, for the outgoing nucleons,
as well as of the two-nucleon induced channel. This strict interplay between
theory and experiments is at the basis of the present belief that the
$\Gamma_n/\,\Gamma_p$ puzzle has been solved. In particular, this is due to the
study of nucleon coincidence observables, recently measured at
KEK \cite{exp5,exp6}, whose weak-decay-model independent analysis carried out
in \cite{th1,th2} yields values of $\Gamma_n/\,\Gamma_p$ around
$0.3\div 0.4$ for the $^5_\Lambda He$ and $^{12}_\Lambda C$ hypernuclei, in
satisfactory agreement with the most recent theoretical
evaluations \cite{L2,L4c,L8,Dq2}. New, more precise results are
expected from forthcoming experiments at DA$\Phi$NE \cite{daphne} and
J-PARC \cite{jparc}.

Another intriguing issue is represented by the asymmetry of the angular
emission of non-mesonic decay protons from polarized hypernuclei.
The large momentum transfer involved in the $n(\pi^+,K^+)\Lambda$ reaction can
be exploited to produce final hypernuclear states characterized by a relevant
amount of spin-polarization, preferentially aligned along the axis normal to
the reaction plane \cite{pol1,pol2}.
The hypernuclear polarization mainly descends from a non-negligible spin-flip
term in the elementary $n\pi^+\to\Lambda K^+$ scattering process, which in turn interferes
with the spin-non-flip contribution \cite{asym}: in free space, and for
$|\p_\pi|=1.05$ GeV and $\theta_K\simeq 15^o$, the final hyperon
polarization is about 75\%.

Polarization observables represent a natural playground to test the present
knowledge of the NMWD reaction mechanism, being strictly
related to the spin-parity structure of the elementary $\Lambda N\to nN$
interaction. Indeed, by focusing on the $p$-induced channel, experiments with
polarized hypernuclei revealed the existence of an asymmetry in the angular
distributions of the emitted protons with respect to the hypernuclear
polarization direction. Such an asymmetry originates from an interference
effect between parity-violating and parity-conserving amplitudes for the
$\overrightarrow{\Lambda}p\to np$ elementary process, and can thus complement
the experimental information on the $\Gamma_n$ and $\Gamma_p$ partial decay
rates, which are instead mainly determined by the parity-conserving
contributions. As for the $\Gamma_n/\,\Gamma_p$ ratio, FSI could play a crucial
role in determining the measured value of this observable.

The asymmetry puzzle concerns the strong disagreement between theoretical
predictions and experimental extractions of the so-called intrinsic asymmetry
parameter $a_\Lambda$.
The first asymmetry measurements \cite{pol1,pol2} with limited statistics
gave large uncertainties and even inconsistent results.
The very recent and more accurate data from KEK-E508 \cite{rev2a,Asexp2,Asexp3}
favour small values of  $a_\Lambda$, compatible with a vanishing value.
Moreover, the observed asymmetry parameters are negative for $^{12}_\Lambda C$
and positive (and smaller, in absolute value) for $^5_\Lambda He$. Theoretical
models generally predict negative and larger values of $a_\Lambda$. FSI effects do not
improve the agreement with
data \cite{GarbarinoAsymmetry}. The inclusion, within the usual framework of
nonrelativistic OME models, of the exchange of correlated and uncorrelated
pion pairs \cite{GarbarinoTPE} greatly improves the situation. Indeed, it only
slightly modifies the non-mesonic decay rates and the $\Gamma_n/\,\Gamma_p$
ratio, but the modification in the strength and sign of some relevant decay
amplitudes is crucial and yields asymmetry parameters which lie well within the
experimental observations. In particular, a small and positive value is now
predicted for $a_\Lambda$ in $^{5}_\Lambda \overrightarrow{He}$. This important
achievement justifies the claim that also the asymmetry puzzle has finally
found a solution.

Recent experimental and theoretical studies have led to a deeper understanding
of some fundamental aspects of the NMWD of $\Lambda$-hypernuclei.
From a theoretical point of view, the standard approach towards these topics
has been strictly nonrelativistic, with both nuclear matter and finite nuclei
calculations converging towards similar conclusions: nonrelativistic full
one-meson-exchange plus two-pion-exchange models, based on the
polarization-propagator method (PPM) \cite{PPM2,PPM3} or on the
wave-function method (WFM) \cite{L4,L2,L4c}, seem able to reproduce all the
relevant observables for the $^5_\Lambda He$ and $^{12}_\Lambda C$
light-medium hypernuclei. A crucial contribution to this achievement is
however due to a non-trivial theoretical analysis of KEK most recent
coincidence data, based on the proper consideration and simulation of nuclear
FSI and two-nucleon induced decays  \cite{th2,GarbarinoTPE,ICC,ICCcorr}.
In those nonrelativistic models many theoretical ingredients are included
with unavoidable approximations.
Initial-state interactions and strong $\Lambda N$ SRC are treated in a
phenomenological way, though based on microscopic models calculations.
The inclusion of the full pseudo-scalar and vector mesons spectra, in
particular the strange $K$ and $K^*$ mesons, in the context of complete OME
nonrelativistic calculations, somewhat clashes with the still poor knowledge
of the weak $\Lambda$-$N$-meson coupling constants for mesons heavier than
the pion. Their evaluation requires model calculations which unavoidably
introduce a certain degree of uncertainty in the corresponding conclusions
about the $\Gamma_n/\Gamma_p$ ratio and other observables.
In order to reproduce few observables, i.e., the decay rates $\Gamma_{nm}$,
$\Gamma_{n}$, and  $\Gamma_{p}$, and the asymmetry parameter $a_\Lambda$ for
$^5_\Lambda He$ and $^{12}_\Lambda C$, these models need to include many
dynamical effects, such as the exchange of all the possible mesons, plus
two-pion exchange, plus phenomenological $\sigma$ mesons, plus the
corresponding interferences, and, moreover, strong nuclear medium effects in
the form of non-trivial FSI. Although the inclusion of many  theoretical
ingredients can be considered as a natural and desirable refinement of the
simple OPE models, all the improvements do not seem to provide a significantly
deeper insight into the decay dynamics.
Despite all the theoretical efforts,
the solution of the $\Gamma_n/\Gamma_p$ and $a_\Lambda$ parameter puzzles seems
due to effects, such as distortion, scattering and absorption of the primary
nucleons by the surrounding nuclear medium, rather than to the
weak-strong interactions driving the elementary $\Lambda N\to nN$ or
$\Lambda NN\to nNN$ process. The re-analysis of the recent KEK
experimental data \cite{exp1,exp5,exp6,E307corr} and the corresponding
extraction of $\Gamma_n/\,\Gamma_p$ are indeed completely independent of the
weak-decay-mechanisms \cite{th2,th5}, but depend strongly on the model adopted
to describe FSI and on somewhat arbitrary assumptions, e.g. on the ratio
$\Gamma_2/\Gamma_1$ between two-nucleon and one-nucleon induced non-mesonic
decay rates.

The presently available experimental information on hypernuclear decay is still
limited and affected by uncertainties of both experimental and theoretical
nature. Moreover, single-nucleon spectra seem to point at a
possible systematic protons underestimation \cite{th5}. The new generation of
experiments planned in various laboratories worldwide is expected to produce
more precise data on the already studied observables as well as new
valuable information in the form of differential energy and angular decay
particles spectra.

In spite of the recent important achievements, the NMWD of hypernuclei deserves
further experimental and theoretical investigation.
From the theoretical point of view, the role of relativity is almost unexplored.
But for a few calculations in \cite{relPS,rel1,rel2,Cmes2} no fully relativistic model
has been exploited to draw definite conclusions about the role of relativity in
the description of the weak decay dynamics.

In this paper we present a fully relativistic model for the NMWD of $^{12}_{\Lambda}C$  \cite{PHD}.
The adopted framework consists of a finite nucleus WFM approach  based on
Dirac phenomenology. As a first step the model includes only OPE and
one-kaon-exchange (OKE) diagrams, and is limited to one-nucleon
induced decay. We are aware that the neglected contributions could play an
important role in the decays.
Our aim is to explain all the at least qualitative features of the hypernuclear
NMWD with a conceptually simple model, in terms of a few physical mechanisms
and free parameters.  We stress that, dealing with a fully relativistic
treatment of the weak dynamics based on the calculation of Feynman diagrams
within a covariant formalism, it is quite difficult to directly compare such an
approach and its results to standard nonrelativistic OME calculations.
We will thus rather focus on the internal coherence and on the theoretical
motivations of the model.

The model is presented in Sec. 2.  Numerical results for the total
non-mesonic decay width $\Gamma_{nm}$, the $\Gamma_n/\Gamma_p$ ratio, the
$a_\Lambda$ intrinsic asymmetry parameter, as well as for kinetic energy and
angular spectra are presented and discussed in Sec. 3. The sensitivity to the
choice of the main theoretical ingredients is investigated. The theoretical
predictions of the model are compared with the most recent experimental
results. Some conclusions are drawn in Sec. 4.


\section{Model}
\label{sec:theory}

In this Section we present a fully relativistic finite nucleus wave-function
approach to study the NMWD of the ${}^{12}_{\,\Lambda}C$ hypernucleus.
Our model is based on a fully relativistic evaluation of the
elementary amplitude for the $\Lambda N\to NN$ process, which, at least in the
impulse approximation, is the fundamental interaction responsible for
the NMWD. Covariant, complex amplitudes are calculated in terms of proper
Feynman diagrams. The tree-level diagram involves a weak and a strong current,
connected by the exchange of a single virtual meson. Integrations over the
spatial positions of the two vertices as well as over the transferred 3-momentum are
performed. As a first approximation, only OPE and OKE diagrams are considered.
Possible two-nucleon induced contributions are neglected, even if they could play an
important role in the hypernuclear decay phenomenology.

Interested readers can find further details about the present model in the PhD thesis
of Ref. \cite{PHD}, where an extensive analysis of the adopted formalism as well as of
the involved theoretical ingredients is provided.

In the calculation of the hypernuclear decay rate the Feynman amplitude must
then be properly included into a many-body treatment for nuclear structure. The
amplitude is therefore only a part of the complete calculation, but it is the
basic ingredient of the model and involves all the relevant information on the
dynamical mechanisms driving the decay process.

Short range correlations are also included in the model,
coherently with what commonly done in most nonrelativistic calculations, since
the relatively high nucleon energies involved in the hypernuclear NMWD can in
principle probe quite small baryon-baryon distances, where strong interactions
may be active and play an important role. Following a phenomenological approach,
we have chosen to include initial SRC effects by means of a multiplicative local and
energy-independent function, whose general form \cite{relPS} provides an
excellent parametrization of a realistic $\Lambda N$ correlation function
obtained from a G-matrix nonrelativistic calculation \cite{Gmatrix,Halderson}.
The problem of ensuring a correct implementation of such a nonrelativistic SRC
function within a relativistic, covariant formalism has been addressed in
Ref. \cite{SRC} and shown to be tightly connected with the choice of the
interaction vertices. For full generality, we also choose to account for possible
strong $NN$ short range interactions acting on the two final emitted nucleons, again
adopting a simple phenomenological average correlation function \cite{L2} which
provides a good description of nucleon pairs in $^4{\rm He}$ \cite{finalSRC} as calculated
with the Reid soft-core interaction \cite{finalSRC2}; such final-state SRC could in principle
play an important role, and they complement the final-state interactions between
each of one of the two emitted nucleons and  the residual nulceus, that is
accounted for in our model by a relativistic complex optical potential. 

\subsection{Coupling ambiguities}
\label{sec:CouplingAmbiguities}

In order to devise a relativistic treatment of the elementary $\Lambda N\to NN$
process, great care must be devoted to the choice of the Dirac-Lorentz structure
for the strong and weak parity-conserving vertices. The pseudo-scalar ($PS$)
prescription, that consists in a $i\gamma_5$ Dirac structure, and the
pseudo-vector ($PV$) one, that contains a $\gamma_5\gamma^{\mu}\partial _{\mu}$
axial-vector structure, are in principle equivalent, at least for positive
energy on-shell states, because they descend from equivalent Lagrangians.
However, ambiguities arise when one tries to take into account SRC in terms of
a multiplicative local and energy-independent function $f(r)$. Such
ambiguities are not of dynamical origin and should not be mistaken as
relativistic effects: they are simply bound to the phenomenological way of
including (initial and final) short range correlations, by matching a nonrelativistic correlation
function within a relativistic Feynman diagram approach.
The crucial observation, in this regard, is that it is possible to give
theoretical reasons \cite{SRC} to prefer the $PV$ coupling in its
modified version where the $4$-derivative operates on the propagator
$(PV^{\prime})$, over the $PS$ coupling and also over the standard $PV$ one,
where the $4$-derivative acts on the matrix element. On the one hand, the
$PV^{\prime}$ choice permits to recover, in the nonrelativistic limit, the
standard OPE potential, multiplied by $f(r)$, which is commonly used as the
starting point in nonrelativistic calculations, whereas the $PS$ and $PV$
couplings yield a simple Yukawa function in the same limit: this allows, at
least in principle, a comparison between relativistic and nonrelativistic results.
On the other hand, a microscopic model of (initial) SRC effects, adopting standard
$PS$ $NN\pi$ vertices and introducing an additional
$\omega$-exchange mechanism simultaneous to the OPE dominant one, produces a
result analogous to what can be derived in a phenomenological tree-level
approach contemplating the inclusion of a SRC function, provided in this case
the modified derivative $PV^{\prime}$ coupling, rather than the $PS$ one, is
employed. The main feature is the development of an explicit dependence of the
interaction matrix elements on the exchanged three-momentum (through the
momentum involved in the corresponding loop integrals, in the microscopic
model, or the derivative effect of the $PV^{\prime}$ coupling, in the
tree-level phenomenological approach). When dealing with a more complex model
for nuclear structure, we do not generally use positive energy on-shell states.
Still the general message keeps its validity, though the details of the
explicit calculations may be different. In order to correctly treat SRC,
nuclear currents showing a dependence on the 3-momentum transfer $\q$ are
needed, which in the simple model above correspond to matrix elements between
external spinors and intermediate spinors carrying the momentum of the
intermediate state excited by the heavy meson. This can be achieved  using
the $PV^{\prime}$ coupling acting on the pion field, while the use of the $PS$ or
standard $PV$ couplings, as done for instance in Ref. \cite{relPS}, would
generate nuclear currents independent of $\q$, corresponding, within the
considered simple SRC model, to matrix elements between spinors all carrying
the external momenta.

\subsection{Pseudo-scalar couplings}
\label{sec:pseudoScalarCoupling}

As a first example, we employ a $PS$ coupling for the strong vertex and for the
parity-conserving part of the weak interaction.
The $\Lambda N \to nN$ fundamental process can then be decomposed into a weak
$\Lambda N \pi$ vertex, governed by the weak Hamiltonian
\be
{\cal H}^{(w)}_{\Lambda N \pi}=i\,G_F\,m_\pi^2\,\bar{\Psi}_N^{(s)}\left(A_\pi+B_\pi\,\g_5\right)\ta\cdot\bm{\phi}_\pi\,\Psi_{\Lambda}^{(b)}\,,
\ee
and a strong $N N \pi$ vertex, driven by the Hamiltonian
\be
{\cal H}^{(s)}_{N N \pi}=i\,g_{NN\pi}\,\bar{\Psi}_N^{(s)}\,\g_5\,\ta\cdot\bm{\phi}_\pi\,\Psi_{N}^{(b)}\,.
\ee
The Dirac spinors $\Psi_{\Lambda}^{(b)}$ and $\Psi_{N}^{(b)}$ are the wave
functions of the bound $\Lambda$ hyperon and nucleon inside the hypernucleus,
$\bar{\Psi}_N^{(s)}$ is the Dirac spinor representing the scattering
wave function of each one of the two final nucleons, $\ta$ is the vector formed
by the three Pauli matrices, and $\bm{\phi}_\pi$ is the isovector pion field.
The Fermi weak constant $G_F$ and the pion mass $m_\pi$ give
$G_F\,m_\pi^2 \simeq 2.21\times10^{-7}$. The empirical constants $A_\pi=1.05$
and $B_\pi=-7.15$ are adjusted to the free $\Lambda$ decay and determine the
strengths of the parity-violating and parity-conserving non-mesonic weak
rates, respectively. Finally, $g_{NN\pi}=13.16$ is the strong $N N \pi$
coupling.
The initial $\Lambda$ and final nucleon fields, $\Psi_{\Lambda}^{(b)}$ and
$\bar{\Psi}_N^{(s)}$,  are  defined in space-spin-isospin space and they are
described by a space-spin part times a two component isospinor. In addition,
the $\Lambda$ field is represented as a pure $m_{t_\Lambda}=-1/2$ state to
enforce the empirical $\Delta I=1/2$ selection rule.

\begin{figure}[htbp]
\begin{center}
\includegraphics[width=12cm]{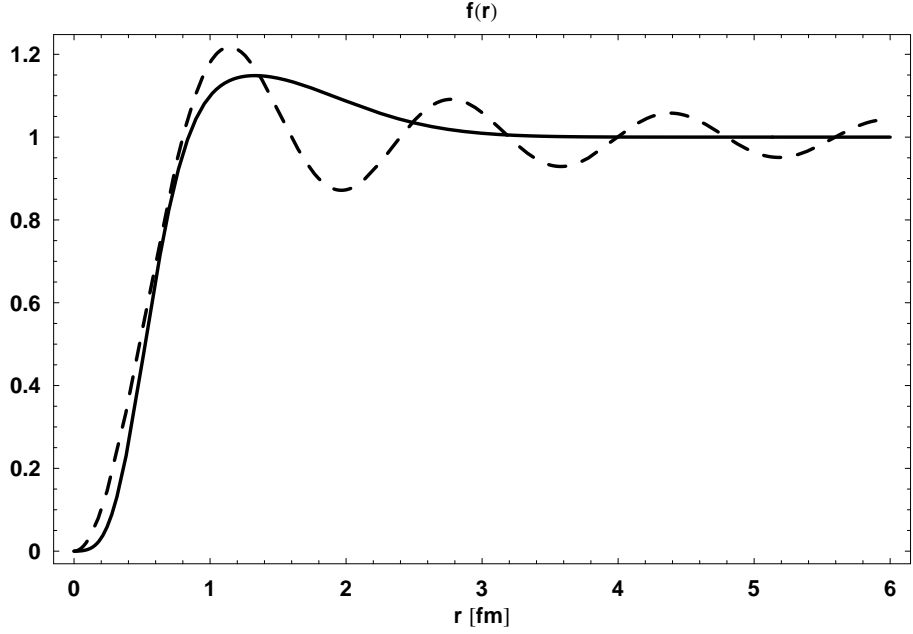}
\caption{\small{Initial $\Lambda N$ $f_{\Lambda N}^{ini}(r)$ 
(Eq. (\ref{SRC1}), solid line) and final $NN$ $f_{NN}^{fin}(r)$ 
(Eq. (\ref{SRC2}), dashed line) correlation functions, with $r$ 
representing the baryon-baryon relative distance.}}
\label{fig:SRC}
\end{center}
\end{figure}

The relativistic Feynman amplitude for the two-body matrix element describing
the $\Lambda N\to NN$ transition, driven by the exchange of a virtual pion,
can be written as
\ba
{\cal T}_{fi,\,\pi}^{(PS)}&=&i\,G_F\,m_\pi^2\,g_{NN\pi}\int {\rm d}^4x \int {\rm d}^4 y\;\,f_{\Lambda N}^{ini}(|\x-\y|)\nn \\
&\times&\Big[\bar{\Psi}^{(s)}_{\k_1,m_{s_1},m_{t_1}}(x)\left(A_\pi+B_\pi\,\g_5\right)
\tau_1^a\,\Psi^{(b)}_{\alpha_\Lambda,\mu_\Lambda,-1/2}(x)\Big]\nn\\[0.1cm]
&\times&\,\,\delta^{ab}\,\Delta_\pi(x-y)\,\,f_{NN}^{fin}(|\x-\y|)\, \nn \\[0.15cm]
&\times&\Big[\bar{\Psi}^{(s)}_{\k_2,m_{s_2},m_{t_2}}(y)\,\g_5\,\tau_2^b\,\Psi^{(b)}_{\alpha_N,\mu_N,m_{t_N}}(y)\Big]\,,
\label{Tfi}
\ea
where $\Psi^{(b)}_{\alpha_\Lambda,\mu_\Lambda,-1/2}$ and
$\Psi^{(b)}_{\alpha_N,\mu_N,m_{t_N}}$ are the bound $\Lambda$ and nucleon wave
functions, with $\alpha_{N,\Lambda}=\{nlj\}_{N,\Lambda}$ quantum numbers and
total spin (isospin) projections $\mu_{N,\Lambda}$ ($m_{t_N}$,
$m_{t_\Lambda}=-1/2$), and $\Psi^{(s)}_{\k_i,m_{s_i},m_{t_i}}$, with $i=1,2$,
are the scattering wave functions
for the two final nucleons emitted in the hypernuclear NMWD, with asymptotic
momenta $\k_i$ and spin (isospin) projections $m_{s_i}$ ($m_{t_i}$).
In both the initial and the final baryon wave functions it is possible to factor
out the isospin 2-spinors as well as the energy-dependent exponentials:
$\Psi(x)\equiv \psi(\bm{\x})\,e^{-i\, E\, x^0}\,\chi_{t=1/2}^{m_t}$, where $E$ is
the total energy of the considered baryon.
The factor $f_{\Lambda N}^{ini}(|\x-\y|)$ represents a short-range two-body
correlation function acting on the initial $\Lambda$ and $N$ baryons, and similarly 
$f_{NN}^{fin}(|\x-\y|)$ describes possible short range interactions between the two final nucleons
 emerging from the interaction vertex. $\Delta_\pi(x-y)$ is the Fourier transform of the
product of the pion propagator with the vertex form factors (supposed to be
equal for the strong and weak vertices), i.e.,
\ba
\Delta_\pi(x-y)=\int\frac{{\rm d}^4
q}{(2\pi)^4}\,\frac{e^{iq\cdot(x-y)}}{q^2-m_\pi^2+i\varepsilon}\;{\cal F}_\pi^{\,2}(q^2) \ . \label{Tfo}
\ea
After performing time integrations in Eq. (\ref{Tfi}) and taking advantage of
the $q^0$ part of the integral in Eq. (\ref{Tfo}), we get for the relativistic amplitude the expression
\ba
{\cal T}_{fi,\,\pi}^{(PS)}&=&i\,G_F\,m_\pi^2\,g_{NN\pi}\,\,{\cal I}\int {\rm d}^3\x \int {\rm d}^3 \y \;\,f_{\Lambda N}^{ini}(|\x-\y|)\nn \\
&\times&\Big[\bar{\psi}^{(s)}_{\k_1,m_{s_1}}(\x)\left(A_\pi+B_\pi\,\g_5\right)\psi^{(b)}_{\alpha_\Lambda,\mu_\Lambda}(\x)\Big] \nn \\[0.1cm]
&\times & \,\,\Delta_\pi(|\x-\y|)\,\,f_{NN}^{fin}(|\x-\y|)\,
\Big[\bar{\psi}^{(s)}_{\k_2,m_{s_2}}(\y)\,
\g_5\,\psi^{(b)}_{\alpha_N,\mu_N}(\y)\Big] \nn \\[0.2cm]
&\times &\,(2\pi)\,\,\delta\left(E_1+E_2-E_\Lambda-E_N\right) \ ,
\label{Tfi2}
\ea
where
\be
\Delta_\pi(| \x - \y |) \equiv \int \frac{\diff^3 \q}{(2\pi)^3}
\frac{e^{-i\q\cdot(\x-\y)}}
{(q^0)^2-\q^2-m_\pi^2+i\varepsilon} \,\,
{\cal F}_\pi^{\,2}\left((q^0)^2-\q^2\right)\,\bigg|_{q^0=\widetilde{q}^{\,0}} \, , \label{Dpion}
\ee
with $\widetilde{q}^{\,0}=E_\Lambda-E_1=E_2-E_N$ and ${\cal I}$  is an isospin factor that depends on the considered decay channel (either $p$- or $n$- induced), i.e.,
\ba
{\cal I}&\equiv&
\left[\left(\chi_{1/2}^{m_{t_1}}\right)^\dagger\ta_1\,\,\chi_{1/2}^{-1/2}\right]\cdot
\left[\left(\chi_{1/2}^{m_{t_2}}\right)^\dagger\ta_2\,\,\chi_{1/2}^{m_{t_N}}\right] \ .\label{isofact}
\ea
It is easy to check that ${\cal I}$ is different from zero only for the charge-conserving processes $\Lambda p\to np$ and $\Lambda n\to nn$.
For the calculation of the integral over the 3-momentum transfer $\q$ in
Eq. (\ref{Tfi2}) we choose a monopolar form factor, i.e.,
\ba
{\cal F}_\pi(q^2)\equiv\frac{\Lambda_\pi^2-m_\pi^2}{\Lambda_\pi^2-q^2}\,,\label{FF}
\ea
where $m_\pi\simeq 140$ MeV is the pion mass and $\Lambda_\pi\simeq 1.3$ GeV
is the cut-off parameter \cite{relPS}.
The initial $\Lambda N$ correlation function adopted in our calculations is \cite{relPS}
\be
f_{\Lambda N}^{ini}(r)=\left(1-e^{-r^2/a^2}\right)^n+b\,r^2 e^{-r^2/c^2}\,,\label{SRC1}
\ee
with $n=2,\,a=0.5,\,b=0.25,\,c=1.28$, while the final $NN$ correlation function is chosen as \cite{L2}
\be
f_{NN}^{fin}(r)=1-j_0(q_c\,r)\,,\label{SRC2}
\ee
where $j_0(x)=\frac{\sin x}{x}$ is the first spherical Bessel function, and $q_c=3.93$ fm$^{-1}$.
The two correlation functions of Eqs. (\ref{SRC1}) and (\ref{SRC2}) are plotted 
in Fig. \ref{fig:SRC}.
As a consequance of the approximations adopted in the present calculation, we 
could then, for practical purposes, treat initial and final SRC as a whole, in 
terms of an overall correlation functions defined as
\be
f(|\x - \y|)=f_{\Lambda N}^{ini}(|\x - \y|)\,f_{NN}^{fin}(|\x - \y|)\,.\label{SRC}
\ee

\subsection{Pseudo-vector couplings}
\label{sec:PseudoVectorCoupling}

When we use derivative $PV^{\prime}$ couplings, the equivalent of
Eq. (\ref{Tfi}) is
\ba
{\cal T}_{fi,\,\pi}^{(PV\,')}&=&
i\,G_F\,m_\pi^2\,g_{NN\pi}\int {\rm d}^4x \int {\rm d}^4y \nn \\
&\times&\,\Big[\bar{\Psi}^{(s)}_{\k_1,m_{s_1},m_{t_1}}(x)
\left(A_\pi-\frac{i B_\pi}{2\bar{M}}\,\g_5\,\g^\mu\,\partial_\mu^x\right)\tau_1^a\,\Psi^{(b)}_{\alpha_\Lambda,\mu_\Lambda,-1/2}(x)\Big]
\nn \\[0.2cm]
&\times&\,\quad\delta^{ab}\,\Delta_\pi(x-y)\,f(|\x-\y|)\, \nn \\[0.15cm]
&\times&\,
\Big[\bar{\Psi}^{(s)}_{\k_2,m_{s_2},m_{t_2}}(y)
\left(-\frac{i}{2M_N}\,\g_5\,\g^\nu\,\partial_\nu^y\right)\tau_2^b\,\Psi^{(b)}_{\alpha_N,\mu_N,m_{t_N}}(y)\Big]
\,,
\label{Tfi3}
\ea
where $\bar{M}=(M_N+M_\Lambda)/2$, with $M_\Lambda=1.1156$ GeV, and now $f(|\x-\y|)$ is defined as in Eq. (\ref{SRC}). 
In Eq. (\ref{Tfi3}) the space-time derivatives act just on the pion propagator $\Delta_\pi(x-y)$
(given in Eq. (\ref{Dpion})) and not on the SRC function $f(|\x-\y|)$, which
is considered as a
phenomenological ingredient entering Eq. (\ref{Tfi3}) in a factorized form.
Thus, the involved derivatives translate into multiplicative terms under the
integral over $\q$. After the time integrations in Eq. (\ref{Tfi3}), we obtain
the final expression of the relativistic amplitude
\ba \label{Tfi4}
{\cal T}_{fi,\,\pi}^{(PV\,')}&=&i\,G_F\,m_\pi^2\,g_{NN\pi}\,\,
{\cal I} \int {\rm d}^3\x \int {\rm d}^3 \y \nn \\ &\times &
\Bigg[ \int \frac{{\rm d}^3 \q}{(2\pi)^3}
\frac{e^{-i\q\cdot(\x-\y)}}{q^2-m_\pi^2+i\varepsilon}\,
{\cal F}_\pi^{\,2}(q^2)\,\,f(|\x-\y|)\nn\\[0.1cm]
&&\quad\times\;\Big[\bar{\psi}^{(s)}_{\k_1,m_{s_1}}(\x)\left(A_\pi+\frac{B_\pi}{2\bar{M}}\,\g_5\,\sla{q}\right)
\psi^{(b)}_{\alpha_\Lambda,\mu_\Lambda}(\x)\Big] \nn \\[0.1cm]
&&\quad\times\;\Big[\bar{\psi}^{(s)}_{\k_2,m_{s_2}}(\y)
\left(-\frac{1}{2M_N}\,\g_5\,\sla{q}\right)\psi^{(b)}_{\alpha_N,\mu_N}(\y)
\Big]\nn\\[0.05cm]
&&\quad\times\;\,(2\pi)\,\,\delta\left(E_1+E_2-E_\Lambda-E_N\right)\Bigg]\ ,
\ea
where ${\cal I}$, ${\cal F}(q^2)$, and $f(|\x-\y|)$ are defined in Eqs. (\ref {isofact}), (\ref{FF}), and (\ref{SRC}), respectively. Eq. (\ref{Tfi4}) must be evaluated at
${q^0=\widetilde{q}^{\,0}\equiv E_\Lambda-E_1=E_2-E_N}$.

The crucial difference between Eq. (\ref{Tfi2}) and Eq. (\ref{Tfi4}) is that
in Eq. (\ref{Tfi4}), obtained adopting derivative $PV^{\prime}$  couplings at
the vertices, the matrix elements between the initial bound and the final
scattering states explicitly depend on the 3-momentum transfer $\q$, while
using $PS$ couplings the matrix elements in Eq. (\ref{Tfi2}) are independent
of $\q$.

Though representing a computational complication, the $\q$-dependence of the
matrix elements is a desirable feature in connection with the problem of
correctly including short range correlations in a fully relativistic formalism 
as the one developed here.
The use of $PS$ vertices, which produces baryonic matrix elements only
depending on the external variables, is not coherent with a simple but
significant model of the physical mechanism behind SRC, based on the
simultaneous exchange of a pion plus one or more heavy mesons.  The correlated
Feynman amplitudes involve box (or more complex) diagrams and one expects that
the interaction matrix elements explicitly depend on the momentum involved in
the corresponding loop integrals. This in turn represents a strong motivation
to consider $PV^{\prime}$ couplings as the most appropriate ones for a
fully relativistic approach to the NMWD of $\Lambda$-hypernuclei, since they
prove able to mimic such a physical effect.

\subsection{Initial- and final-state wave functions}

The main theoretical ingredients entering the relativistic amplitudes of
Eqs. (\ref{Tfi2}) and (\ref{Tfi4}) are the vertices operators and
the initial and final baryon wave functions. Since we adopt a covariant
description for the strong and weak interaction operators, the involved wave
functions are required to be 4-spinors. Their explicit expressions are obtained
within the framework of Dirac phenomenology in presence of scalar and vector
relativistic potentials. In the calculations presented in this work the bound
nucleon states are taken as self-consistent Dirac-Hartree solutions derived
within a relativistic mean field approach, employing a relativistic Lagrangian
containing $\sigma$, $\omega$, and $\rho$ mesons contributions
\cite{boundwf,Serot,adfx,lala,sha}.
Slight modifications also permit to adapt such an approach to the determination
of the initial $\Lambda$ wave function and binding energy.
The explicit form of the bound-state wave functions reads
\be\psi_{n\kappa}^{\,\mu}(\bm{r})=\left(
                            \begin{array}{c}
                              g_{n\kappa}(|\bm{r}|)\,\,{\cal Y}_{\kappa}^{\,\mu}(\Omega) \\[0.3cm]
                              i\,f_{n\kappa}(|\bm{r}|)\,\,{\cal Y}_{-\kappa}^{\,\mu}(\Omega) \\
                            \end{array}
                          \right),
\ee
where the 2-components spin-orbital is written as
\be{\cal Y}_{\kappa}^{\,\mu}(\Omega)\equiv\hspace{-0.1cm}\sum_{\mu_l,\,\mu_s=\pm 1/2}\left(\,l\;\,\mu_l\;\,1/2\;\,\mu_s\,\Big|\,j\;\,\mu\,\right)
Y_l^{\mu_l}(\Omega)\,\chi_{1/2}^{\,\mu_s}\,,\ee
with
\be j=|\kappa|-\frac{1}{2}\qquad{\rm and}\qquad\left\{\begin{array}{l}
                                                        \displaystyle{l=\kappa} \hspace{1.65cm}{\rm if}\quad \kappa>0\\[0.2cm]
                                                        \displaystyle{l=-\,\kappa-1} \qquad{\rm if}\quad \kappa<0
                                                      \end{array}
\right.\;;\ee
$n$ is the radial quantum number and $\kappa$ determines both the total and
the  orbital angular momentum quantum numbers. The normalization of the radial
wave functions is given by
\be\int {\rm d}|\bm{r}| \;|\bm{r}|^2\,\Big({\big|\,f_{n\kappa}(|\bm{r}|)\big|}^2+{\big|\,g_{n\kappa}(|\bm{r}|)\big|}^2\Big)=1\,.\ee

The outgoing nucleons wave functions are calculated by means of the
relativistic energy-dependent complex optical potentials of Ref. \cite{chc},
which fits proton elastic-scattering data on several nuclei in an energy range up to 1040 MeV.
In the explicit construction of the ejectile states, the direct Pauli reduction
method is followed. It is well known that a Dirac 4-spinor, commonly
represented in terms of its two Pauli 2-spinor components
\be\psi_{\,\bm{k},m_s}(\bm{r})=\left(
                  \begin{array}{c}
                    \phi_{\,\bm{k},m_s}(\bm{r}) \\
                    \chi_{\,\bm{k},m_s}(\bm{r}) \\
                  \end{array}
                \right),
\ee
can be written in terms of its positive energy component $\phi(\bm{r})$ as
\be\psi_{\,\bm{k},m_s}(\bm{r})=\left(
                  \begin{array}{c}
                  \phi_{\,\bm{k},m_s}(\bm{r}) \\[0.15cm]
                  \displaystyle{\left[\frac{\left(\bm{\sigma}\cdot\bm{k}\,\right)}{M_N+E+S(|\bm{r}|)-V(|\bm{r}|)}\right]\phi_{\,\bm{k},m_s}(\bm{r})} \\
                  \end{array}
                \right),\label{Paulired}\ee
where $S(|\bm{r}|)$ and $V(|\bm{r}|)$ are the scalar and vector potentials for the final nucleon with energy $E$. The upper component $\phi(\bm{r})$ can be related to a Schr\"odinger-like wave function $\widetilde{\phi}(\bm{r})$ by the Darwin factor $D(|\bm{r}|)$,
i.e.,
\be \phi(\bm{r})\equiv \sqrt{D(|\bm{r}|)}\,\,\widetilde{\phi}(\bm{r})\,,\ee
with
\be D(|\bm{r}|)\equiv\frac{M_N+E+S(|\bm{r}|)-V(|\bm{r}|)}{M_N+E}\,.\label{Darwin}\ee
The two-component wave function $\widetilde{\phi}(\bm{r})$ is solution of a
Schr\"odinger equation containing equivalent central and spin-orbit potentials,
which are functions of the energy-dependent relativistic scalar and vector
potentials $S$ and $V$. Its general form is given by
\ba\widetilde{\phi}_{\,\bm{k},m_s}(\bm{r})&\hspace{-0.25cm}=\hspace{-0.25cm}&\sqrt{\frac{M_N+E}{2E}}\,\,\sum_{l\,m_l\,j\,\mu}\,4\pi\,i^l\,
\left[u_{lj}(|\bm{r}|)\,\,{\cal Y}_{lj;\,|\bm{k}|}^\mu(\Omega_{\bm{r}})\right]\nn\\
&&\times\,\,\Big(\,l\;m_l\;1/2\;m_s\,\Big|\,j\;\mu\,\Big)\,\,Y_l^{m_l\,*}(\Omega_{\bm{k}})\,.\ea

\subsection{Decay rates}
\label{sec:DecayRates}
In the complete calculation of the total and partial decay rates, as well as of
polarization observables, the dynamical information on the elementary
$\Lambda N\to NN$ process, given by the amplitudes in Eq. (\ref{Tfi2}) or
Eq. (\ref{Tfi4}), are included in a many-body calculation for nuclear structure.
The weak non-mesonic total decay rate is defined as \cite{relPS,NR1}
\ba
\Gamma_{nm}&=&\int\frac{d^3\k_1}{(2\pi)^3}\int\frac{d^3\k_2}{(2\pi)^3}\;(2\pi)\,\delta\left(M_H-E_R-E_1-E_2\right)\nn \\ [0.1cm]
&&\times\,\,
\frac{M_N^2}{E_1E_2}\,\frac{1}{2J_H+1}\hspace{-0.2cm}
\sum_{\begin{array}{c}
                       \vspace{-0.15cm}\scriptstyle{ M_{J_H}\, \{R\}}\\
                       \vspace{-0.15cm}\scriptstyle{ m_{s_1}\, m_{s_2}}\\
                       \scriptstyle{ m_{t_1}\, m_{t_2}}
                      \end{array}}
                      \hspace{-0.3cm}
{\left|{\cal M}_{fi}\right|}^2\, .\label{Gamma}
\ea
The energy-conserving delta function connects the sum of the asymptotic
energies $E_{1,2}$ of the two outgoing nucleons, coming from the underlying
$\Lambda N\to NN$ microscopic process, with the difference between the initial
hypernucleus mass $M_H$ and the total energy $E_R$ of the residual
$(A-2)$-particle system after the decay. A sum over $E_R$ is also usually
understood. Integration over the phase spaces of the two
final nucleons is needed, since the decay rate is a fully inclusive observable.
Moreover, the sums in Eq. (\ref{Gamma}) encode an average over the initial
hypernucleus spin projections $M_{J_H}$, where $J_H$ is the hypernucleus total
spin, a sum over all the spin and isospin quantum numbers of the
residual $(A-2)$-system, $\{R\}\equiv\{J_R,M_R,T_R,M_{T_R}\}$, as well as a sum
over the spin and isospin projections of the two outgoing nucleons
$m_{s_{1,2}}$ and $m_{t_{1,2}}$, respectively.
If we choose  a reference frame in which, for instance, the $\hat{\bm{z}}$-axis
is aligned along the momentum $\k_1$, and exploiting the energy-conservation in
the delta function, the six-dimensional integral in Eq. \ref{Gamma} can be
reduced to a two-dimensional integral, one over the energy of one of the two
final nucleons and the other one over the relative angle between the momenta of
the two nucleons (due to azimuthal symmetry), which can be performed
numerically.

The expression for the NMWD rate  $\Gamma_{nm}$ can be decomposed into a sum over $n$- and
$p$-induced decay processes without any interference effects, i.e.,
\ba\Gamma_{nm}=\sum_{m_{t_N}}\Gamma_{nm}[m_{t_N}]
=\Gamma_{nm}^{(p)}+\Gamma_{nm}^{(n)}\,,\label{Gamma2}
\ea
where $\Gamma_{nm}[m_{t_N}]$ is defined as in Eq. (\ref{Gamma}) and
$|{\cal M}_{fi}|^2$ is evaluated with a fixed value of the initial-nucleon
isospin projection, $m_{t_N} = 1/2$  for $p$-induced and $m_{t_N} = -1/2$ for
$n$-induced channels. Actually, in each term of
$\sum_{m_{t_1}\,m_{t_2}}|{\cal M}_{fi}|^2$ the $m_{t_{1,2}}$ quantum numbers
are fixed, so that $|{\cal M}_{fi}|^2$ would involve products of the kind
${\cal T}_{fi,\,\pi}^{A*}[m_{t_1},m_{t_2},m'_{t_N}]\,{\cal T}_{fi,\,\pi}^{A}[m_{t_1},m_{t_2},m_{t_N}]$,
where, in principle, also interference  effects, $m'_{t_N}\neq m_{t_N}$, are
allowed.
However, the non diagonal products with $m'_{t_N}\neq m_{t_N}$ are
necessarily zero, since if one of the two amplitudes is non-zero the other one
must vanish as a consequence of the  charge-conservation isospin factor
$\cal{I}$ of Eq. (\ref{isofact}) (same final state but different initial states,
$p\Lambda$ or $n\Lambda$). Therefore, only the diagonal terms,
$m'_{t_N}= m_{t_N}$, contribute and without interferences the coherent sum
over $m_{t_N}$ becomes an incoherent one.

The nuclear transition amplitude, from the initial hypernuclear state to the
final state of an $(A-2)$ residual nucleus and the two outgoing nucleons, is
defined as
\ba
{\cal M}_{fi}=\langle \,f\,|\,\hat{{\cal M}}_{\Lambda N\to NN}\,|\,i\,\rangle\,
\ea
and can be represented in terms of the elementary two-body $\Lambda N\to NN$
relativistic Feynman amplitude, ${\cal T}_{fi,\,\pi}$, which contains
all the relevant information about the weak-strong dynamics driving the global
decay process.
The final $A$-particle state $|\,f\,\rangle$ must be further specified and
decomposed into products of antisymmetric two-nucleon and residual
$(A-2)$-nucleon wave functions.
An explicit decomposition for the initial hypernuclear
wave function $|\,i\,\rangle$ can be developed following the approach introduced
in Ref. \cite{relPS}, which is based on a
weak-coupling scheme, i.e., the isoscalar $\Lambda$ hyperon is assumed to be in
the $1s_{1/2}$ ground state and it  only couples to the ground-state wave
function of the $(A-1)$-nucleon core. As discussed in Ref. \cite{relPS}, this
weak-coupling approximation has been able to yield quite good results in hypernuclear
shell-model calculations \cite{weakcoupl}.

The final expression for ${\cal M}_{fi}$ is
\ba
{\left|{\cal M}_{fi}\right|}^2[m_{t_N}]
&=&{\left(\,{T}_R\, {M}_{T_R}\, 1/2\; m_{t_N}\,\Big|\,T_H\,M_{T_H}\right)}^2\, \nn\\[0.15cm]
&\times&\sum_{j_N}
\;A\;{\left\langle\,J_c\,T_H\,\Big\{\Big|\,{J}_R\,{T}_R\,,
j_N\,m_{t_N}\right\rangle}^2\nn\\
&\times&\Bigg[\,\sum_{m_N}{\left({J}_R\, {M}_R\, j_N\, m_N\,\Big|\,J_c\,M_c\right)}^2
\nn\\
&&\quad\times\;
{\left(J_c\, M_c\, 1/2\; m_\Lambda\,\Big|\,J_H\, M_{J_H}\right)}^2\,\,
{\left|{\cal T}_{fi,\,\pi}^A\right|}^2\Bigg],\qquad\quad\label{Mfi2}
\ea
where $m_{t_N}=+1/2$ for $\Gamma_{nm}^{(p)}$ and  $-1/2$ for
$\Gamma_{nm}^{(n)}$, $\{J_H,M_{J_H},T_H,M_{T_H}\}$ are the spin-isospin quantum
numbers for the initial hypernucleus, $\{j_N,m_N\}$ are the initial-nucleon
total spin and its third component, $\{J_c,M_c\}$ are the same quantum numbers
for the $(A-1)$-nucleon core, and, finally,
$m_\Lambda$ is the initial $\Lambda$ total spin projection. In Eq. (\ref{Mfi2})
$\langle J_c\,T_H\,\{|\,\widetilde{J}_R\,\widetilde{T}_R\,, j_N\rangle$ are the
real coefficients of fractional parentage (c.f.p.), which allow the
decomposition of the initial $(A-1)$-nucleon core wave functions in terms of
states involving a single nucleon coupled to a residual $(A-2)$-nucleon
state. The factor $A$ is produced by the combination of  initial- and
final-state antisymmetrization factors with the number of $\Lambda N$ pairs
contributing to the total decay rate. Eq. (\ref{Mfi2}) neglects possible
quantum interference effects between different values of $j_N$ (and $m_N$),
namely we are ruling out interferences between different shells ($s_{1/2}$ and
$p_{3/2}$) for the initial nucleon. Thus the calculation does not
require the c.f.p., but only the spectroscopic factors $S=A\,(c.f.p)^2$, that
can be taken, e.g. from Ref. \cite{relPS}.

\subsection{Antisymmetrization and isospin factors}
\label{sec:Antisymmetrization}

A crucial role in determining the $\Gamma_n/\Gamma_p$ ratio is played by the
isospin content of the model, namely the ${\cal I}$ factors defined in
Eq. (9) in terms of the SU(2) isospin operators (generally
represented by the $2\times 2$ Pauli matrices) and of the corresponding
isospin 2-spinors for the initial $\Lambda$ and $N$ as well as for the two
final nucleons.

Taking advantage of the $\Delta I=1/2$ isospin selection rule, from the
isospin point of view, the $\Lambda$ behaves like a neutron state. We can then
explicitly represent the isospin spinors for the $p$,  $n$ and $\Lambda$ baryons as
\be\chi_p\equiv\chi_{1/2}^{m_{t_N}=1/2}=\left(
                                          \begin{array}{c}
                                            1 \\
                                            0 \\
                                          \end{array}
                                        \right),\qquad
\chi_n\equiv\chi_{1/2}^{m_{t_N}=-1/2}=\left(
                                          \begin{array}{c}
                                            0 \\
                                            1 \\
                                          \end{array}
                                        \right),\qquad
\chi_\Lambda=\chi_n\, .
\ee
With these definitions, the ${\cal I}$   isospin factors can be evaluated for
all the possible combinations of the $m_{t_N}$, $m_{t_1}$, and $m_{t_2}$ isospin
projection quantum numbers. They are non-zero only for those processes in
which charge is conserved, namely $\Lambda p\to np$ and $\Lambda n\to nn$. We  obtain
\bea&&\hspace {-0.5cm}{\cal I}\,\left[m_{t_N}=1/2,\,m_{t_1}=-1/2,\,m_{t_2}=1/2\right]
\,\equiv\,{\cal I}_{\Lambda p\to np}^{\,(d)}\,=\,-1\,,\\
&&\hspace {-0.5cm}{\cal I}\,\left[m_{t_N}=1/2,\,m_{t_1}=1/2,\,m_{t_2}=-1/2\right]\,
\equiv\,{\cal I}_{\Lambda p\to np}^{\,(e)}\,=\,2\,,\\
&&\hspace {-0.5cm}{\cal I}\,\left[m_{t_N}=-1/2,\,m_{t_1}=-1/2,\,m_{t_2}=-1/2\right]
\,\equiv\,{\cal I}_{\Lambda n\to nn}^{\,(d)}\,=\,
{\cal I}_{\Lambda n\to nn}^{\,(e)}\,=\,1\, .
\qquad\quad\eea
All the others possibilities imply charge violation and give zero. The $(d)$
and $(e)$ apices  refer to the direct and exchange diagrams
of the relativistic Feynman amplitudes for the elementary processes.

The use of the isospin formalism means that we are treating the neutron and the proton
as two indistinguishable particles;  therefore
the final state is composed of two identical particles and this requires the
antisymmetrization
 of the ${\cal T}_{fi,\,\pi}^{A}$ amplitude.
 The antisymmetrization acts on the two final nucleons,
exchanging their spin-isospin quantum numbers, $m_{s_i},\,m_{t_i}$, and their
momenta $\bm{k}_i$ within the matrix elements defining the ${\cal T}_{fi,\,\pi}$
complex amplitude. We can thus define
\be{\cal T}_{fi,\,\pi}^{A}\equiv{\cal T}_{fi,\,\pi}^{(d)}-{\cal T}_{fi,\,\pi}^{(e)}\,,\ee
where ${\cal T}_{fi,\,\pi}^{(d)}$ is the Feynman amplitude for the direct diagram,
given by Eqs. (5) or (12),
while ${\cal T}_{fi,\,\pi}^{(e)}$ represents the Feynman amplitude for the exchange diagram,
obtained from the same Eqs. (5) or (12), but with the interchanges
$m_{s_1}\leftrightarrow\, m_{s_2}$, $m_{t_1}\leftrightarrow\, m_{t_2}$
and $\bm{k}_1\leftrightarrow\,\bm{k}_2$. In addition,
the antisymmetrization involves different ${\cal I}$ factors
for the direct and exchange diagrams:
${\cal I}_{\Lambda p\to np}^{\,(d)}\,=\,-1$ and
${\cal I}_{\Lambda p\to np}^{\,(e)}\,=\,2$ for a final $np$ pair, and
${\cal I}_{\Lambda n\to nn}^{\,(d,e)}\,=\,1$ for a final $nn$ pair.
Taking advantage of the factorization
${\cal T}_{fi,\,\pi}^{(d,e)}\equiv{\cal I}^{(d,e)}\,
\widetilde{{\cal T}}_{fi,\,\pi}^{(d,e)}$, the antisymmetrized Feynman amplitudes
can be written as
\bea
\Lambda p\to np\,:\hspace{2.1cm} {\cal T}_{fi,\,\pi}^A&=&{\cal I}_{\Lambda p\to np}^{(d)}\,\widetilde{{\cal T}}_{fi,\,\pi}^{(d)}
-{\cal I}_{\Lambda p\to np}^{(e)}\,\widetilde{{\cal T}}_{fi,\,\pi}^{(e)}\nn\\[0.15cm]
&=&(-)\left[\,\widetilde{{\cal T}}_{fi,\,\pi}^{(d)}+2\,\widetilde{{\cal T}}_{fi,\,\pi}^{(e)}\,\right]\,,\label{pA}\\[0.4cm]
\Lambda n\to nn\,:\hspace{2.1cm} {\cal T}_{fi,\,\pi}^A&=&{\cal I}_{\Lambda n\to nn}^{(d)}\,\widetilde{{\cal T}}_{fi,\,\pi}^{(d)}
-{\cal I}_{\Lambda n\to nn}^{(e)}\,\widetilde{{\cal T}}_{fi,\,\pi}^{(e)}\nn\\[0.15cm]
&=&\left[\,\widetilde{{\cal T}}_{fi,\,\pi}^{(d)}-\widetilde{{\cal T}}_{fi,\,\pi}^{(e)}\,\right]\,.\label{nA}\eea
We stress that many complex amplitudes with different quantum numbers contribute to the
calculation of the nuclear transition amplitude. It is therefore difficult to
make simple estimates of the final result.

\subsection{Asymmetries in polarized hypernuclei decay}
\label{sec:AsymmetriesInPolarizedHypernucleiDecay}

The main effect that is obtained with polarized hypernuclei
is given by the angular asymmetry in the distribution of the emitted protons
with respect to the direction of the hypernuclear polarization. It can be
shown \cite{relPS} that the non-mesonic partial decay rate for the
proton-induced $\overrightarrow{\Lambda} p\to np$ process can be written as
\ba
\Gamma_{nm}^{(p)}=\frac{1}{2J_H+1}\sum_{M_{J_H}}\sigma(J_H,M_{J_H})\equiv I_0(J_H)\, ,\label{sigma}
\ea
where $\sigma(J_H,M_{J_H})\equiv\sum_f{|\langle\,f\,|\,\hat{{\cal M}}\,|
\,i\,;J_H,M_{J_H}\,\rangle|}^2$ is the intensity of protons emitted along the
quantization axis $z$ for a spin projection $M_{J_H}$ of the hypernuclear total spin $J_H$.
In terms of the isotropic intensity for an unpolarized hypernucleus, $I_0(J_H)$, the intensity
of protons emitted in the non-mesonic decay of a polarized hypernucleus (through
the $\overrightarrow{\Lambda} p \to np$
elementary process) along a direction forming an angle $\Theta$ with the polarization axis is defined by
\ba I(\Theta,J_H)=I_0(J_H)\,\Big[1+P_y(J_H)\,A_y(J_H)\,\cos\Theta\Big]\,,\ea
where $P_y(J_H)$ is the hypernuclear polarization and $A_y(J_H)$ the hypernuclear asymmetry parameter,
both depending on the specific hypernucleus under consideration. The asymmetry
$A_y(J_H)$ is a property of the non-mesonic decay and it only depends on
the dynamical mechanism driving the weak decay. In contrast, $P_y(J_H)$ also
depends on the kinematical and dynamical features of the associated production
reaction. The explicit expression for $A_y(J_H)$ reads
\ba A_y(J_H)\equiv\frac{3}{J_H+1}\,\frac{\sum_{M_{J_H}}\sigma(J_H,M_{J_H})\,\,M_{J_H}}
{\sum_{M_{J_H}}\sigma(J_H,M_{J_H})}\,,\label{Asymm}\ea
in terms of the quantities $\sigma(J_H,M_{J_H})$ defined in Eq. (\ref{sigma}).

Within the framework of the shell-model weak coupling scheme, supposing that
the $\Lambda$ hyperon sits in the $1s$ orbital and interacts (weakly) only with
the nuclear core ground-state, angular momentum algebra can be employed to
relate the polarization $p_\Lambda$ of the $\Lambda$ spin inside the
hypernucleus to the hypernuclear polarization $P_y$
\be
p_\Lambda(J_H)=\left\{\begin{array}{l}
\displaystyle{-\,\frac{J_H}{J_H+1}\,P_y(J_H)\,,\qquad{\rm if}\quad J_H=J_c-\frac{1}{2}}\,,\\[0.4cm]
\displaystyle{P_y(J_H)\,,\hspace{2.27cm}{\rm if}\quad J_H=J_c+\frac{1}{2}}\,,
         \end{array}
\right.\label{wcs1}\ee
where $J_c$ denotes the total spin of the $(A-1)$-nucleon core. It turns out useful to introduce an
intrinsic $\Lambda$ asymmetry parameter, $a_\Lambda$, which should be independent of the
considered hypernucleus, such that
\be P_y(J_H)\,A_y(J_H)=p_\Lambda(J_H)\,a_\Lambda\,.\label{wcs2}\ee
The $a_\Lambda$ parameter removes the dependence on the hypernuclear spin $J_H$
and is thus given by
\be
a_\Lambda=\left\{\begin{array}{l}
\displaystyle{-\,\frac{J_H+1}{J_H}\,A_y(J_H)\,,\qquad{\rm if}\quad J_H=J_c-\frac{1}{2}}\,,\\[0.4cm]
\displaystyle{A_y(J_H)\,,\hspace{2.27cm}{\rm if}\quad J_H=J_c+\frac{1}{2}}\,,
         \end{array}
\right.\label{wcs3}\ee
and $a_\Lambda=0$ for $J_H=0$.
Therefore, in this weak-coupling picture, which is known to provide a good approximation for
describing the ground state of $\Lambda$-hypernuclei and is particularly reliable for the
non-mesonic decay (where nuclear structure details are not so important), $a_\Lambda$ can be
interpreted as the intrinsic $\Lambda$ asymmetry parameter for the elementary reaction
$\overrightarrow{\Lambda}p\to np$, involving the polarized hyperon inside the hypernucleus,
and should no more depend on the particular hypernucleus considered.

\subsection{$K$-exchange}
\label{sec:KExchange}

The exchange of the pseudo-scalar $K$ iso-doublets is believed to represent the dominant dynamical mechanism beyond
the simple OPE, since the $K$ meson is the lightest one ($m_K\simeq 495$ MeV) after the pion. It is thus
useful to investigate the role of $K$-exchange in our relativistic model.
Actually, such a contribution was included in all the most recent
nonrelativistic calculations, and in turn proved useful to improve the theory-experiment matching
concerning  the $\Gamma_n/\,\Gamma_p$ decay ratio. The OKE process is driven
by the two following Hamiltonians
\cite{NR1}:
\ba
{\cal H}^{(w)}_{NNK}&\hspace{-0.25cm}=\hspace{-0.25cm}& i\,G_F\,m_\pi^2 \left[
\bar{\Psi}_N^{(s)}\,\chi_{1/2}^{-1/2}\left(C_K^{pv}+C_K^{pc}\,\g_5\right)\left(\phi_K
\right)^\dagger\Psi_{N}^{(b)}\right.\nn\\
&&+\, \left. \bar{\Psi}_N^{(s)}\,\Psi_{N}^{(b)}\left(D_K^{pv}+D_K^{pc}\,\g_5\right)
\left(\phi_K\right)^\dagger\chi_{1/2}^{-1/2}\right],\ea
for the weak (strangeness-changing) $NNK$ vertex, and
\be{\cal H}^{(s)}_{\Lambda N K}=i\,g_{\Lambda NK}\,\bar{\Psi}_N^{(s)}\,\g_5\,\,\phi_K\,
\Psi_{\Lambda}^{(b)}\,,\ee
for the strong (strangeness-conserving) $\Lambda NK$ one. In the previous expressions,
$\phi_K$ is the $K$ meson field and \ba
\chi_{1/2}^{-1/2}=
\left(\begin{array}{c} 0 \\ 1 \end{array}\right) \label{spu}
\ea
is introduced as usual to enforce the $\Delta I=1/2$ isospin rule.
We use the Nijmegen value $g_{\Lambda NK}=12.0$ for the strong $\Lambda N K$ coupling
\cite{nijmegen}, and the weak parity-violating and parity-conserving coupling constants
are
$ C_K^{pv}=0.76,\  C_K^{pc}=-18.9,\  D_K^{pv}=2.09,\  D_K^{pc}=6.63$ \cite{L6}.
Using the  $PS$ form for the strong and weak (parity-conserving) vertices and
proceeding as in the case of OPE, we obtain the analog of
Eq. (\ref{Tfi2}) for the OKE relativistic Feynman amplitude
\ba
{\cal T}_{fi,\,K}^{(PS)}
&=&i\,G_F\,m_\pi^2\,g_{\Lambda NK}\,(2\pi)\,\,\delta\left(E_1+E_2-E_\Lambda-E_N\right)
\nn\\
&\times& \Bigg\{\,{\cal I}\int {\rm d}^3\x \int {\rm d}^3 \y\;\,
\Big[\bar{\psi}^{(s)}_{\k_1,m_{s_1}}(\x)\,\g_5\,\psi^{(b)}_{\alpha_\Lambda,\mu_\Lambda}
(\x)\Big] \nn\\[0.1cm]
&&\quad\times\;\; \Delta_K(|\x-\y|)\,f(|\x-\y|)\nn\\[0.25cm]
&&\quad\times\; \Big[\bar{\psi}^{(s)}_{\k_2,m_{s_2}}(\y)
\left(\frac{C_K^{pv}}{2}+\frac{C_K^{pc}}{2}\,\,\g_5\right)\psi^{(b)}_{\alpha_N,\mu_N}
(\y)\Big]\nn\\[0.2cm]
&&\hspace{0.4cm}+\,\;{\cal K}\int {\rm d}^3\x \int {\rm d}^3 \y\;\,
\Big[\bar{\psi}^{(s)}_{\k_1,m_{s_1}}(\x)\,\g_5\,\psi^{(b)}_{\alpha_\Lambda,\mu_\Lambda}
(\x)\Big]\nn\\[0.2cm]
&&\quad\;\hspace{0.35cm}\times\;\; \Delta_K(|\x-\y |)\,f(|\x-\y| ) \nn \\[0.25cm]
&&\quad\;\hspace{0.35cm}\times\; \Big[\bar{\psi}^{(s)}_{\k_2,m_{s_2}}(\y)
\left\{\left(\frac{C_K^{pv}}{2}+D_K^{pv}\right)\right. \nn \\[0.1cm]
&&\qquad\quad\;\hspace{0.4cm}+\left.
\left(\frac{C_K^{pc}}{2}+D_K^{pc}\right)\g_5\right\}\psi^{(b)}_{\alpha_N,\mu_N}
(\y)\Big]\,\Bigg\},
\label{TfiK}
\ea
where the overall short range correlation function $f(|\x-\y|)$ is given in Eq. (\ref{SRC}) 
and the $K$ propagator has the same structure as in Eq. (\ref{Dpion}),
with a monopolar form factor ${\cal F}_K(q^2)$ for the baryon-baryon-$K$
vertices (the same for the weak and strong vertices),
with $\Lambda_K=1.2$ GeV \cite{NR1}.
The isospin factor ${\cal I}$ is the same that already enters the OPE amplitude
and it is given in Eq. (\ref{isofact}), while the isospin factor $\cal{K}$
is defined as
\ba{\cal K}\,\equiv\,\left[\left(\chi_{1/2}^{m_{t_1}}\right)^\dagger\,
\chi_{1/2}^{-1/2}\right]
\left[\left(\chi_{1/2}^{m_{t_2}}\right)^\dagger\,\chi_{1/2}^{m_{t_N}}\right]\,.\ea
As before
the {\cal K} isospin factors are non-zero only for those processes in
which charge is conserved, namely $\Lambda p\to np$ and $\Lambda n\to nn$.
The quantities $\cal{K}$ are:
\bea
&&\hspace {-0.5cm}{\cal K}\,\left[m_{t_N}=1/2,\,m_{t_1}=-1/2,\,m_{t_2}=1/2\right]\,\equiv\,
{\cal K}_{\Lambda p\to np}^{\,(d)}\,=\,1\,,\\
&&\hspace {-0.5cm}{\cal K}\,\left[m_{t_N}=1/2,\,m_{t_1}=1/2,\,m_{t_2}=-1/2\right]\,\equiv\,
{\cal K}_{\Lambda p\to np}^{\,(e)}\,=\,0\,,\\
&&\hspace {-0.5cm}{\cal K}\,\left[m_{t_N}=-1/2,\,m_{t_1}=-1/2,\,m_{t_2}=-1/2\right]\,\equiv\,
{\cal K}_{\Lambda n\to nn}^{\,(d)}\,=\,
{\cal K}_{\Lambda n\to nn}^{\,(e)}\,=\,1\, .\qquad\quad\eea
The final-state antisymmetrization requires to consider an antisymmetrized Feynman
amplitude, defined  as the difference of the direct and exchange diagram contributions,
where the space-spin part of the exchange term is obtained from
Eqs. (36) and (38) by interchanging the momenta $\bm{k}_{1,2}$ and spin quantum numbers
$m_{s_{1,2}}$ of the outgoing nucleons scattering wave functions.
The complete expressions can be found in Ref. (54).

When we consider derivative $PV^{\prime}$ couplings for the strong and
weak parity-conserving interactions,
the OKE Feynman amplitude is given by (see Eq. (\ref{Tfi4}))
\ba
{\cal T}_{fi,\,K}^{(PV\,')}
&=&i\,G_F\,m_\pi^2\,g_{\Lambda NK}\,(2\pi)\,\,\delta\left(E_1+E_2-E_\Lambda-E_N\right)\nn\\
&\times&
\Bigg\{\,{\cal I}\int {\rm d}^3\x \int {\rm d}^3 \y \int\frac{{\rm d}^3\q}{(2\pi)^3}\;
\frac{e^{-i\q\cdot(\x-\y)}}{q^2-m_\pi^2+i\varepsilon}\nn \\[0.15cm]
&&\quad\times\;\; {\cal F}_K^{\,2}\left(q^2\right)f(|\x-\y|)\nn\\[0.2cm]
&&\quad\times\; \Big[\bar{\psi}^{(s)}_{\k_1,m_{s_1}}(\x)\left(-\frac{1}{2\bar{M}}\,\,\g_5\,\sla{q}\right)
\psi^{(b)}_{\alpha_\Lambda,\mu_\Lambda}(\x)\Big]\nn\\[0.1cm]
&&\quad\times\; \Big[\bar{\psi}^{(s)}_{\k_2,m_{s_2}}(\y)\Bigg(\frac{C_K^{pv}}{2}+
\frac{C_K^{pc}}{4M_N}\,\,\g_5\,\sla{q}\Bigg)\psi^{(b)}_{\alpha_N,\mu_N}
(\y)\Big]\nn\\[0.2cm]
&&\hspace{0.4cm}+\,\; {\cal K}\int {\rm d}^3\x \int {\rm d}^3 \y \int\frac{{\rm d}^3\q}{(2\pi)^3}\;
\frac{e^{-i\q\cdot(\x-\y)}}{q^2-m_\pi^2+i\varepsilon}\nn\\[0.2cm]
&&\quad\;\hspace{0.35cm}\times\;\; {\cal F}_K^{\,2}\left(q^2\right)f(|\x-\y|)\nn\\[0.2cm]
&&\quad\;\hspace{0.35cm}\times\; \Big[\bar{\psi}^{(s)}_{\k_1,m_{s_1}}(\x)\left(-\frac{1}{2\bar{M}}\,
\g_5\,\sla{q}\right)\psi^{(b)}_{\alpha_\Lambda,\mu_\Lambda}(\x)\Big] \nn\\[0.15cm]
&&\quad\;\hspace{0.35cm}\times\; \Big[\bar{\psi}^{(s)}_{\k_2,m_{s_2}}(\y)
\left\{\left(\frac{C_K^{pv}}{2}+D_K^{pv}\right)\right.\nn\\[0.1cm]
&&\left.\qquad\quad\;\hspace{0.35cm}+\left(\frac{C_K^{pc}}{4M_N}+
\frac{D_K^{pc}}{2M_N}\right) \g_5\,\sla{q}\right\}
\psi^{(b)}_{\alpha_N,\mu_N}(\y)\Big]\,\,\Bigg\}\,,\label{TfiK2}
\ea
where again $q^0$ must be fixed at $\widetilde{q}^0\equiv E_\Lambda-E_1=E_2-E_N$.
As in the OPE case, the use of $PV^{\prime}$ couplings, with derivative terms
acting on the $K$ propagator, gives an expression for the amplitude in which the
$\q$-integral cannot be simply factorized, as instead happens when adopting
$PS$ vertices. The isospin factors are non-zero only for those processes in which charge is
conserved, i.e., $\Lambda p\to np$ and $\Lambda n\to nn$. When considering the
$(\pi+K)$ combined effect, the non-mesonic decay rate becomes proportional to
the squared modulus of the (antisymmetrized) sum of the OPE and OKE
relativistic Feynman amplitudes, i.e.,
\ba
\Gamma_{nm} \propto
{\left|\,\left[{\cal T}_{fi,\,\pi}^{(PS,PV\,')}\right]^A+
\left[{\cal T}_{fi,\,K}^{(PS,PV\,')}\right]^A\,\right|}^2 \,,
\ea
thus the two contributions add coherently and $\pi$-$K$ interference effects
may arise.

\section{Results and discussion}
\label{sec:results}

In this Section the main results of our model for the
$^{12}_\Lambda C$ non-mesonic decay are presented. We compare our relativistic
finite-nucleus calculation with the results of the KEK experiments performed
during the last five years.

Numerical results obtained with different choices for some important
theoretical ingredients are compared to point out and investigate the role and
relevance of each ingredient to the final results. Both $PS$ and $PV{^\prime}$
prescriptions for the vertex structure are analyzed, even if there are
theoretical reasons to prefer  the modified $PV^{\prime}$ prescription, i.e.,
with derivatives acting on the propagator of the exchanged meson, as already
discussed in Sec. \ref{sec:theory}.
The decay dynamics must be considered with great care. As a first step, we have
included the OPE diagram, that is the simplest  contribution to the process and
is supposed to represent the bulk of the weak decay, and the OKE diagram, that
is  the most natural refinement of the model.
The role of SRC is also discussed. The standard choice for many calculations is given by
Eq. (\ref {SRC1}), that is a multiplicative function that represents a
parametrization of a realistic $\Lambda N$ correlation function obtained in
the framework of a many-body calculation. We are aware that this choice can be
suitable only for a nonrelativistic calculation and that its use in a
relativistic calculation is not justified by rigorous theoretical
considerations, but we adopt it as a useful starting point.
A detailed analysis of the role of SRC is anyway beyond the scope of
this paper. An analogous SRC function, sharing all the basic features of the 
$\Lambda N$ initial one (locality, energy-independence, multiplicative action) 
can also be introduced to model short-range correlations between the two emitted 
nucleons (see Eq. (\ref{SRC2})). 

\subsection{Integrated observables}
\label{subsec:int}

In this Section we present our results for
the total non-mesonic decay width $\Gamma_{nm}=\Gamma_n+\Gamma_p$, the
$\Gamma_n/\Gamma_p$ ratio, and the $a_\Lambda$ intrinsic asymmetry parameter.
The numerical results for the OPE  and $\pi+K$ models are
displayed in Table \ref{tab:rispi} and \ref{tab:riskaon}, respectively. 
As anticipated, we explore three possible choices for the
short range baryon-baryon correlation functions: we can include both initial- and 
final-state SRC, i.e. we set $f_{\Lambda N}^{ini}(r)$ as in Eq. (\ref{SRC1}) 
and $f_{NN}^{fin}(r)$ as in Eq. (\ref{SRC2}); or we take into account only 
initial $\Lambda N$ correlations, i.e., we set $f_{\Lambda N}^{ini}(r)$ again 
as in Eq. (\ref{SRC1}) and $f_{NN}^{fin}(r)=1$; or we completely switch off 
SRC, i.e., we set  $f_{\Lambda N}^{ini}(r)=f_{NN}^{fin}(r)=1$.
We notice that the configuration in which only final SRC are active, while the
initial ones are switched off, is quite unnatural and not of particular 
interest: therefore, it will not be considered here.

In Table \ref{tab:risex} the most recent experimental results for $\Gamma_{nm}$,
$\Gamma_n/\,\Gamma_p$, and $a_\Lambda$ are given to provide a reference scheme.

We note that in our approach we neglect the two-nucleon induced decay channel,
that can give a contribution of about 20-25\% to $\Gamma_{nm}$
\cite{daphne,jparc,kim09}.

\begin{itemize}
	\item \textbf{$\pi$-exchange}: 	As a first step, we have considered
the OPE contribution, that is the starting point of all OME models, and
we have evaluated the amplitude for both  $PS$ and $PV\,'$ couplings
with the three possible SRC choices. The results are shown in Table \ref{tab:rispi}.
The main difference between the $PS$ and $PV{^\prime}$ cases is that the use of
$PS$ vertices yields total decay rates $\Gamma_{nm}$ considerably higher than those
obtained with $PV{^\prime}$ vertices: in the $PS$ case $\Gamma_{nm}$ is more than two
times larger than  in the $PV{^\prime}$ case and, moreover, it overestimates the
experimental result.
The $\Gamma_n/\,\Gamma_p$ ratio is less affected by the vertex choice:
the ratio calculated with $PV{^\prime}$ couplings, and including initial and final SRC, is larger
than the corresponding one with $PS$ couplings by about 25\%, but remains within the experimental range.
The asymmetry is negative in both cases, when SRC are accounted for: negative 
and small with $PS$ couplings ($-0.080$ or $-0.064$, depending on our choice to 
include or not final SRC besides initial ones) and larger in absolute value 
with $PV{^\prime}$ couplings ($-0.126$ and $-0.108$, respectively).
The effects of final SRC in a calculations where initial SRC are included, is
generally small, independently of the vertices choice: the total decay rate is
increased by no more than 5\% and even smaller effects are obtained on the 
neutron-to-proton ratio. More significant changes (20-30\%) are found for the 
asymmetry that, being close to 0, is more sensitive to the various
contributions.
If we compare the results with and without SRC, we see that without SRC, in the 
$PS$ sector,
$\Gamma_{nm}$ increases by about 25$\%$, 
$\Gamma_n/\,\Gamma_p$ increases by about 20$\%$, and $a_\Lambda$, which with 
SRC is small and negative, changes its sign, becoming positive and still 
remaining small in size.
Similar results are obtained when the role of SRC is analyzed with 
$PV{^\prime}$ couplings: neglecting SRC, $\Gamma_{nm}$ and  
$\Gamma_n/\,\Gamma_p$ are increased, with respect to the results 
obtained in presence of 
SRC, by about 15\% and 10\%, respectively, and $a_\Lambda$ increases and 
approaches 0, still remaining negative.

\item \textbf{($\pi$ $+$ $K$)-exchange}: The inclusion of the OKE contribution
has significant, in some cases even sizeable effects on the integrated results
shown in  Table \ref{tab:riskaon}. The decay rate $\Gamma_{nm}$ is somewhat
reduced, more significantly for $PV{^\prime}$ couplings (about 10\%) than for $PS$ ones.
The $PS$ choice overestimates the experimental total non-mesonic decay rate, while the $PV{^\prime}$ choice is in fair
agreement with the experiments, only slightly underestimating (when including SRC effects) the most recent results on $\Gamma_{nm}$.
We note that some underestimation is anyway expected in the present model, where the two-nucleon induced
decay is neglected. The OKE mechanism produces a much larger effect on the $\Gamma_n/\,\Gamma_p$ ratio, whose value is strongly enhanced, by a factor of about 2, in the $PS$ case and reduced, by about 40$\%$, in the $PV{^\prime}$ case.
The difference between the values of $\Gamma_n/\,\Gamma_p$ calculated with $PS$ and $PV{^\prime}$ couplings is therefore strongly enhanced. With $PS$ couplings the ratio becomes now greater than $0.6$, independently of our choice for SRC, in strong disagreement with the most recent experimental extractions based on coincidence spectra. We have already
discussed, however, why we believe that the $PS$ choice is not very reliable. Adopting $PV{^\prime}$ couplings the reduction
produced by the OKE mechanism is helpful to achieve a better agreement with the most recent experimental extraction of
the $\Gamma_n/\,\Gamma_p$ ratio, even though already the OPE results turned out to be satisfactory.
We must notice, however, that in our calculation we get closer to the experimental range from above,
while in most nonrelativistic calculations, where the OPE results are quite small and the inclusion of the OKE mechanism increases the ratio, the same values are approached from below.
The $a_\Lambda$ parameter is somewhat reduced (in absolute value) with the $PS$ choice,
while with the $PV{^\prime}$ choice it is significantly enhanced by the OKE mechanism, also turning from negative
to positive in absence of SRC, slightly worsening the agreement with the experimental result (although both values are within the  experimental range). Also in this case the inclusion of OKE increases the difference between the results with
the $PS$ and $PV{^\prime}$ couplings. All these considerations are generally unchanged by SRC,
which are accounted for in the present model by a simple phenomenological 
correlation function. In general, the effects of SRC are
similar to those obtained in the OPE model and this is independent of the
vertices. The inclusion of final $NN$ SRC, in addition to initial $\Lambda N$
ones, only slightly increases the total decay rate (by just a few percent) while
practically leaving the $\Gamma_n/\Gamma_p$ ratio unchanged; a bigger effect can
be seen on the asymmetry, in particular for $PV{^\prime}$ vertices, even though the very small values of this observable makes it more difficult to draw conclusions on the possible role of the various theoretical ingredients in determining its final size.
When SRC are completely neglected, in the $PV{^\prime}$ case $\Gamma_{nm}$ and $\Gamma_n/\,\Gamma_p$ are enhanced and $a_\Lambda$  increases, up to $0.134$. In the $PS$ case $\Gamma_{nm}$ and $\Gamma_n/\,\Gamma_p$ are again enhanced and $a_\Lambda$ approaches $0$, still keeping the negative sign.

\end{itemize}
\begin{table}
\caption{\label{tab:rispi}\small{Model results for
$\Gamma_{nm}=\Gamma_n+\Gamma_p$,
$\Gamma_n/\,\Gamma_p$, and $a_\Lambda$ for
the $^{12}_\Lambda C$ hypernucleus when only the (OPE) ($\pi$) diagram is considered.}}
\vspace{0.5cm}
\begin{tabular}{||c| c c c ||} \hline
{Model configuration}&
\hspace{0.25cm}{$\Gamma_{nm}/\,\Gamma_\Lambda^{free}$}\hspace{0.25cm} &
$\hspace{0.1cm}\Gamma_n/\,\Gamma_p\hspace{0.1cm}$ & $\hspace{.6cm}a_\Lambda\hspace{.6cm}$\\
\hline &&&\\[-0.25cm]
Pseudo-Scalar ($PS$) couplings &&&\\
\cline{1-1}&&&\\[-0.2cm]
 $\pi$  & 2.426 & 0.342 & $-$0.080 \\[1ex]
 $\pi$, no final SRC   & 2.351 & 0.344 & $-$0.064 \\[1ex]
 $\pi$, no SRC  & 2.950 & 0.413 & 0.052  \\[1ex]
\hline &&&\\[-0.2cm]
Pseudo-Vector ($PV\,'$) couplings &&&\\
\cline{1-1}&&&\\[-0.2cm]
 $\pi$ & 0.995 & 0.436 & $-$0.126 \\[1ex]
 $\pi$, no final SRC & 0.965 & 0.430 & $-$0.108 \\[1ex]
 $\pi$, no SRC  & 1.119 & 0.469 & $-$0.029 \\[0.05cm]
\hline
\end{tabular}
\end{table}

\begin{table}
\caption{\label{tab:riskaon}\small{Model results for
$\Gamma_{nm}=\Gamma_n+\Gamma_p$,
$\Gamma_n/\,\Gamma_p$,
and $a_\Lambda$ for the $^{12}_\Lambda C$ hypernucleus when both OPE and OKE
($\pi+K$) diagrams are considered.}}
\vspace{0.5cm}
\begin{tabular}{||c| c c c ||} \hline
{Model configuration}&
\hspace{0.25cm}{$\Gamma_{nm}/\,\Gamma_\Lambda^{free}$}\hspace{0.25cm} &
$\hspace{0.1cm}\Gamma_n/\,\Gamma_p\hspace{0.1cm}$ & $\hspace{.6cm}a_\Lambda\hspace{.6cm}$\\
\hline &&&\\[-0.25cm]
Pseudo-Scalar ($PS$) couplings &&&\\
\cline{1-1}&&&\\[-0.2cm]
$\pi+K$  & 2.386 & 0.678 & $-$0.060 \\[1ex]
$\pi+K$, no final SRC & 2.277 & 0.677 & $-$0.054 \\[1ex]
$\pi+K$, no SRC  & 2.847 & 0.781 & 0.002 \\[1ex]
 \hline &&&\\[-0.2cm]
Pseudo-Vector ($PV\,'$) couplings &&&\\
\cline{1-1}&&&\\[-0.2cm]
$\pi+K$  & 0.888 & 0.299 & $-$0.038 \\[1ex]
$\pi+K$, no final SRC & 0.863 & 0.299 & $-$0.014 \\[1ex]
$\pi+K$, no SRC  & 1.080 & 0.330 & 0.134 \\[0.05cm]
\hline
\end{tabular}
\end{table}

\begin{table}
\caption{\small{Recent experimental results for $\Gamma_{nm}=\Gamma_n+\Gamma_p$,
$\Gamma_n/\,\Gamma_p$ and $a_\Lambda$ for the $^{12}_\Lambda C$ hypernucleus.}}
\label{tab:risex}
\vspace{0.5cm}
\begin{tabular}{||c| c c c ||} \hline
{Exp.}&
\hspace{0.1cm}{$\Gamma_{nm}/\,\Gamma_\Lambda^{free}$}\hspace{0.1cm} &
$\hspace{0.1cm}\Gamma_n/\,\Gamma_p\hspace{0.1cm}$ & $\hspace{.2cm}a_\Lambda\hspace{.2cm}$\\
\hline &&&\\[-0.2cm]
KEK 2004 (E307)\cite{E307corr} & $0.828\pm 0.087$ & $0.87\pm 0.23$ & \\
 & & $(0.60\pm 0.23)$ &  \\[1ex]
KEK 2003 (E369) \cite{exp1} & & $0.51\pm0.15$ & \\[1ex]
KEK 2004 (E508) \cite{okadames,exp2,Asexp3} & $0.953\pm 0.032$ & $0.5\div0.6$ & $-0.16^{+0.33}_{-0.28}$ \\
(single-nucleon spectra)  & & & \\[1ex]
KEK 2004 (E508) \cite{exp5,exp6} & & $0.51\pm0.14$ & \\
(coincidence spectra) & & & \\[1ex]
KEK 2004 (E508) \cite{exp5,exp6} & & $0.38\pm0.14$ & \\
(coincidence spectra, \cite{th2})  & & $(0.29\pm0.14)$ & \\[1ex]
KEK 2004 (E508) \cite{exp2} & & $0.88\pm0.16$ & \\
(single-nucleon spectra, \cite{th5}) & & $(0.95\pm0.21)$ & \\[1ex]
Exp KEK 2004 (E508) \cite{exp5,exp6} & & $0.46\pm0.09$ & \\
(coincidence spectra, \cite{th5}) & & $(0.43\pm0.10)$ & \\
\hline
\end{tabular}
\end{table}


Our OPE results are different from the results produced by nonrelativistic
models. Usual nonrelativistic calculations with OPE give small values of the  $\Gamma_n/\,\Gamma_p$ ratio, in the range $0.05\div 0.2$, and large negative values of the asymmetry parameter $a_\Lambda$. Our relativistic OPE calculation with
$PV{^\prime}$ couplings, and including initial and final SRC, gives $\Gamma_n/\,\Gamma_p =0.436$ and $a_\Lambda=-0.126$. The addition of OKE pulls the value of $\Gamma_n/\,\Gamma_p$ down to 0.299 and $a_\Lambda$ becomes $-0.038$. Both OPE and OPE+OKE results are in satisfactory agreement with the most recent determinations of the ratio, based on the analyses of coincidence spectra in KEK experiments
(see Table \ref{tab:risex}).

Our calculation gives large values for the ratio also when using $PS$ vertices
($\Gamma_n/\,\Gamma_p =0.342$ with OPE and 0.678 with $\pi+K$-exchange, in presence of initial and final SRC), in
contrast with the results of the relativistic calculation with $PS$ couplings
of Ref. \cite{relPS} ($\Gamma_n/\,\Gamma_p =0.14$ with OPE and 0.25 with
$\pi+K$-exchange). Different results are also obtained concerning the role
of (initial) SRC, whose effect in the calculation of Ref. \cite{relPS} is to give a strong reduction of the
$\Gamma_{nm}$ decay width (about a factor of 4), while in our model a much more moderate reduction is found (and the further inclusion of final-state SRC does not change this picture).
Our model is under many aspects similar to the relativistic one of \cite{relPS}, and initial SRC are described in the two calculations
by the same correlation function (Eq. {\ref{SRC1}}). The differences between the results of the two relativistic calculations are not clear and deserve further investigation.

The general message that can be extracted from various nonrelativistic calculations is that a good agreement with experiments can be reached only by considering the full OME potential and many other effects, like final state Nucleon-Nucleon interactions, rescattering, and intranuclear cascade, while OPE results are always far from the experimental values.
In contrast, our calculations seem to point out a different result and suggest the relative relevance of the OPE and OKE contributions to get closer to the experimental results.
It is not easy to explain the origin of this effect within our fully relativistic model. Actually, unlike what happens with the well known
nonrelativistic potentials, where the contributions from various space-spin-isospin decay channels can be clearly isolated and separately
studied, with our model, where we adopt a completely different approach based on the calculation of relativistic Feynman diagrams, a similar
analysis is not possible, or, at least, this cannot be done in a simple way.

It is not easy to establish a criterion to compare relativistic and
nonrelativistic models and explain their different results. It would be
desirable to perform a careful comparison between our relativistic description
of the current densities and the nonrelativistic operators.
We could start from our $4\times 4 $ Dirac operators and reduce them to
effective $2\times 2$ nonrelativistic operators. However, we believe that this
would give a better understanding of our relativistic calculation and of its
global internal coherence, but not a nonrelativistic reduction to be compared
with the corresponding results of nonrelativistic finite-nuclei calculations
available in the literature.

Besides final $NN$ short-range correlations (mainly acting in proximity of the elementary interaction vertex), in our model only the effects of FSI due to the interaction of each one of the two outgoing nucleons with the residual nucleus are included, through a relativistic
energy-dependent complex optical potential fitted to elastic proton-nucleus
scattering data. 
The use of the same phenomenological optical potentials to
describe FSI has been very successful in describing exclusive
$(e,e^{\prime}p)$, $(\gamma,p)$ data and neutrino-nucleus
scattering \cite{book,Kel,Ud1,Ud3,Ud4,meucci1,meucci2,meucci3,mecpv,meucci4,ee,cc,Meucci:2005ir,Meucci:2006ir}.
The imaginary part of the optical potential produces an absorption of flux, that
is correct for an exclusive reaction, but it would be incorrect for an inclusive
reaction, where all the channels contribute and the total flux must be
conserved. An approach where FSI are treated in inclusive reactions by means of
a complex optical potential and the total flux is conserved is presented
in \cite{ee,cc}. In the present model the hypernuclear non-mesonic weak decay is
treated as a semi-inclusive process where the two outogoing nucleons are detected
and most of the reaction channels that are responsible for the imaginary part of
the optical potential do not contribute.
Rescattering effects where the outgoing nucleons interact with other nucleons in
their way out of the nucleus and
generate secondary nucleons can also affect the decay width. These processes,
which  are not included in the present calculations, are accounted for in
\cite{th5,GarbarinoAsymmetry} by the intranuclear cascade model of
\cite{ICC,ICCcorr}.

In the simplest approach, FSI are neglected and the plane-wave limit is considered for the scattering wave functions. In the
plane-wave limit, calculations with $PV{^\prime}$ vertices give
$ \Gamma_n+\Gamma_p=2.428$, $\Gamma_n/\,\Gamma_p=0.556$, and $a_\Lambda=-0.210$
with only OPE, and $\Gamma_n+\Gamma_p=2.263$, $\Gamma_n/\,\Gamma_p=0.375$, and
$a_\Lambda=-0.031$ with $\pi+K$ exchange.
It is clear that, as it was expected, $\Gamma_{nm}$ is much higher than in
presence of a complex absorptive optical potential.
The $\Gamma_n/\,\Gamma_p$ ratio also increases  with respect to the results
obtained including FSI effects. This is mainly due to the fact that, though
basically isospin-independent, these optical potentials also include Coulomb
correction terms that distinguish between protons and neutrons. Due to the relatively
small energies of the process, these terms are comparable to the central and
spin-orbit Schroedinger-equivalent potentials and they can thus play an
important role within the FSI implementation. The asymmetry parameter is
somewhat reduced in both cases in the plane-wave limit.

\subsection{Kinetic energy spectra}\label{sec:enspec}

\begin{figure}[htbp]
\begin{center}
\includegraphics[width=12cm]{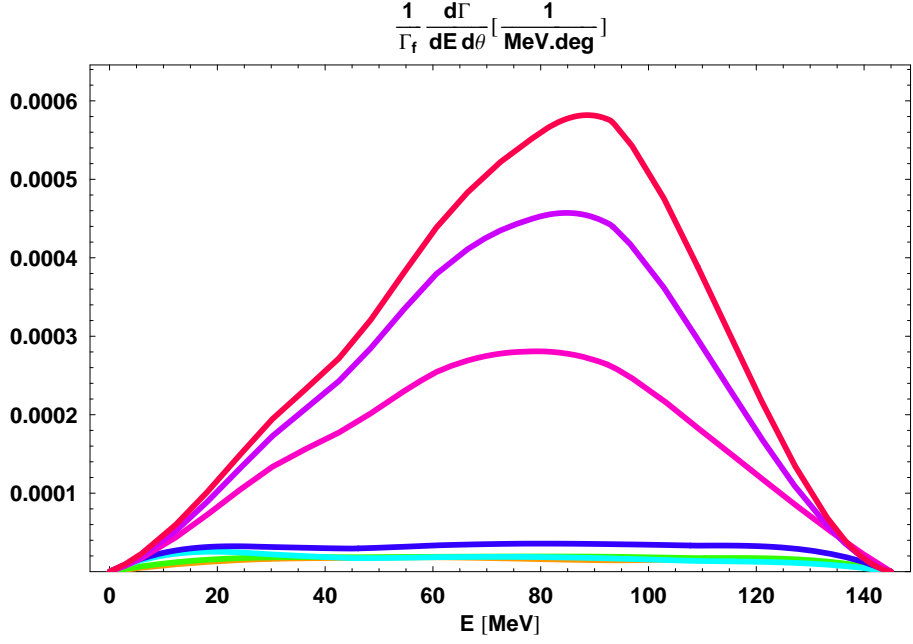}
\caption{\small{Double-differential spectra (normalized to
$\Gamma_\Lambda^{free}$) for the $\Lambda p\to np$ decay channel as a function
of the kinetic energy of the outgoing proton, $T_p \equiv E$, for different
values of the relative angle $\theta$ between the momenta of the two outgoing nucleons.
Line convention: $\theta=30^{o}$ (orange), $\theta=60^{o}$ (green),
$\theta=90^{o}$ (cyan), $\theta=120^{o}$ (blue), $\theta=150^{o}$
(magenta), $\theta=160^{o}$ (purple),  $\theta=170^{o}$ (red).
Calculations are performed with $PV{^\prime}$  couplings and include $\pi+K$ contributions as well as initial and final SRC.}}

\label{fig:Eparz}
\end{center}
\end{figure}

\begin{figure}[htbp]
\begin{center}
\includegraphics[width=12cm]{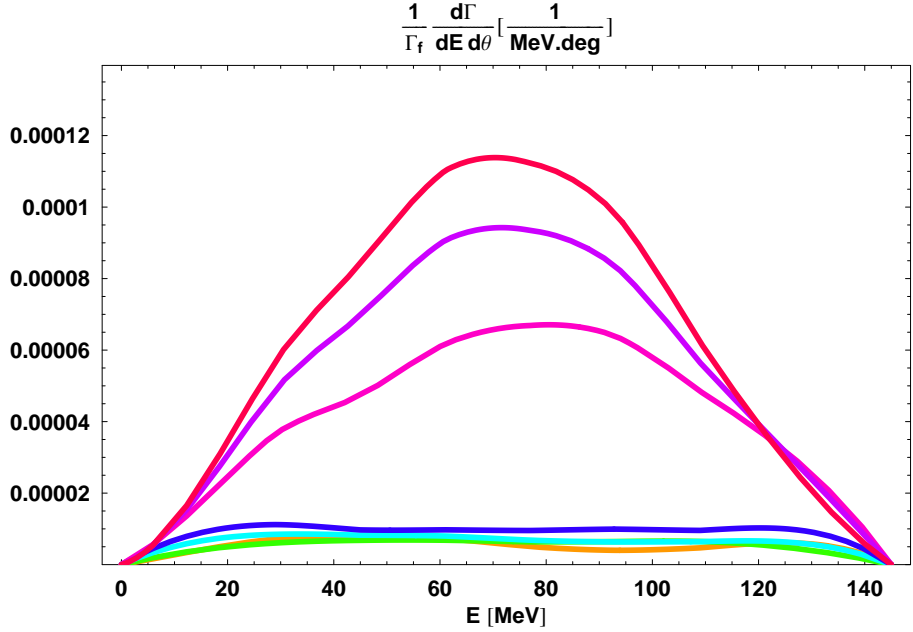}
\caption{\small{The same as in Fig. \ref{fig:Eparz}, but for
neutron kinetic energy spectra, $T_n \equiv E$, of the $\Lambda n\to nn$ decay
channel. }}
\label{fig:EparzN}
\end{center}
\end{figure}

Further insight into the theoretical content of the model is provided by the
investigation of the calculated kinetic energy spectra.
Some examples  are presented and discussed in this Section.
In Figs. \ref{fig:Eparz}-\ref{fig:EparzN} the double-differential spectra,
${\rm d}\Gamma_{p,n}/{\rm d}E\,{\rm d}\theta$, for the $\Lambda p\to np$ and
$\Lambda n\to nn$ decay channels at different angles, are displayed as a function
of the proton  and neutron kinetic energy  $T_{p,n} \equiv E$.
Here $\theta$ is the relative angle between the momenta of the two outgoing nucleons.
The energy spectra are calculated with $PV{^\prime}$ vertices, including both OPE and
OKE contributions as well as initial and final SRC.
It is evident from the results shown in the figures that the energy spectra for
angles $\theta\leq 120^{o}$ display a quite flat behaviour over the whole
possible energy range (from $0$ to $\sim 150$ MeV) and have
comparable sizes, while  the curves obtained for $\theta\geq 120^{o}$, i.e.,
nearly back-to-back angles, are clearly peaked
and increase rapidly in size with $\theta$.
Furthermore, these high-angle distributions suggest the presence of an
underlying double-peak structure, though not very pronounced, with a first small
peak at $E\simeq 30\div40$ MeV and a much larger peak at about
$E\simeq 70\div 90$ MeV. Such a double-peak behaviour can be understood in
terms of the contributions coming from the s$_{1/2}$ and
p$_{3/2}$ shells for the initial proton, which tend to produce different energy
distributions. In the case of proton emission, the spectra are non-symmetric with
respect to $E=T_{max}/2$, where $T_{max}$ is the available kinetic energy for
the emitted nucleons, due to the distorting energy-dependent
optical potential acting on the final nucleons, which distinguishes between
proton and neutron states. On the contrary, in the case of neutron-induced
decay, the symmetry of the problem, which involves the emission of two
indistinguishable neutrons, leads to almost symmetric spectra.


\begin{figure}[htbp]
\begin{center}
\includegraphics[width=12cm]{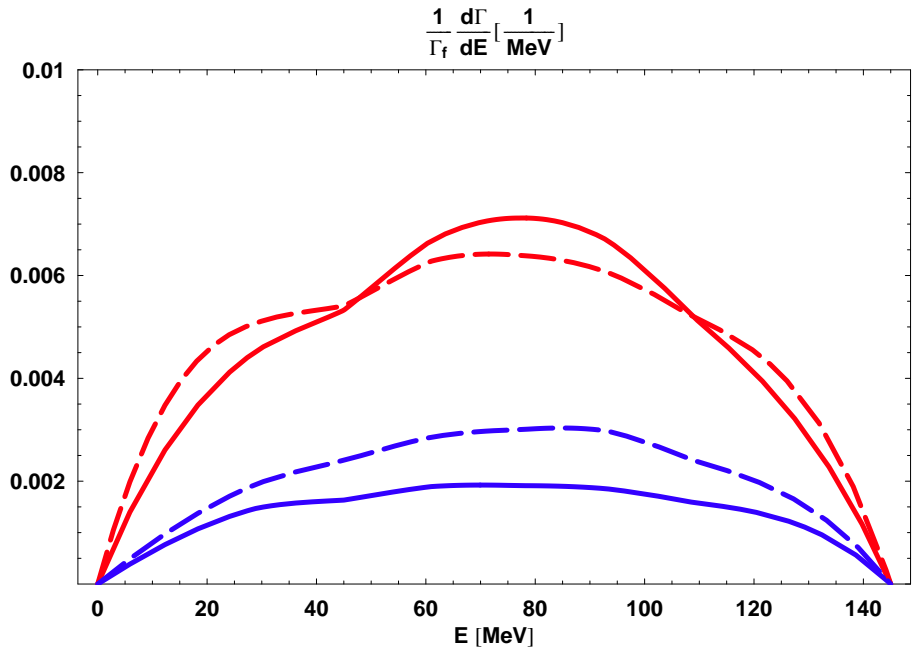}
\caption{\small{Kinetic energy spectra (normalized to $\Gamma_\Lambda^{free}$)
integrated over the angle $\theta$. Calculations are performed with
$PV{^\prime}$ couplings and include both initial and final SRC. Red (blue) lines refer to proton (neutron) spectra, as
obtained in proton-(neutron-)induced one-body decay channels
($E\equiv T_{p,n}$). Dashed lines correspond to results obtained with only
OPE, solid lines are obtained including also OKE contributions.}}
\label{fig:EtotPV}
\end{center}
\end{figure}

\begin{figure}[htbp]
\begin{center}
\includegraphics[width=12cm]{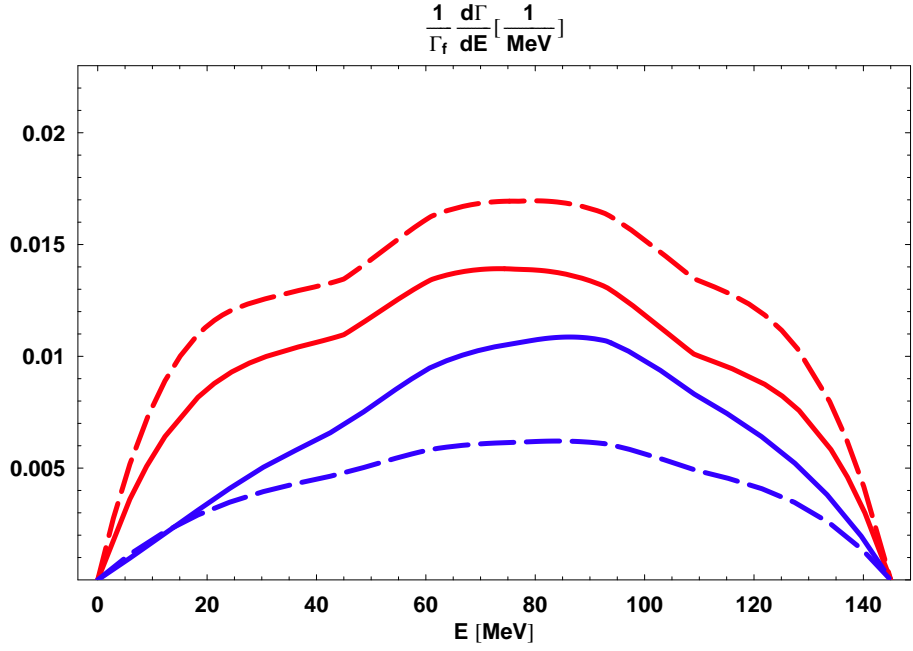}
\caption{\small{The same in Fig. \ref{fig:EtotPV}, but for $PS$ couplings.}}
\label{fig:EtotPS}
\end{center}
\end{figure}

The kinetic energy spectra ${\rm d}\Gamma_{p,n}/{\rm d}E$, for the proton and
neutron-induced decay channels, integrated over the relative angle $\theta$, are
presented in Fig. \ref{fig:EtotPV}. The results obtained with only
OPE and with $\pi+K$ exchange are compared in the figure. The $PV{^\prime}$ choice
for the vertices has been adopted in the calculations, and both initial and final SRC are included.
In the OPE case the proton spectrum is significantly larger in size than the
neutron one. Moreover, the two-peak structure is evident in the proton spectrum, while the shape of the neutron spectrum is smoother
and much more symmetric with respect to $T_{max}/2$. The inclusion of the
OKE contribution  produces a more peaked proton spectrum but does not
modify its global size. In contrast, the neutron spectrum is considerably
reduced by OKE and its shape is slightly flattened. At the level of integrated observables, such a behaviour produces the
reduction of the $\Gamma_n/\,\Gamma_p$ ratio, when including kaon-exchange.
In Fig. \ref{fig:EtotPS} the kinetic energy spectra integrated over $\theta$ are
given for $PS$ couplings.  The proton spectrum with only OPE again shows
an asymmetric shape, with a slight two-peak behaviour (its global scale is now
about twice the one for the corresponding $PV{^\prime}$ case). The
inclusion of the OKE contribution yields an almost identical shape but
reduces the global size of the proton spectrum. The neutron energy spectrum for the
OPE case is smoother and smaller in size with respect to the proton one. When we include the OKE contribution, the neutron spectrum
becomes much more peaked and its size is increased. These results are in
opposite trend with respect to what found for $PV{^\prime}$ couplings in Fig. \ref{fig:EtotPV}
and, by consequence, in the $PS$ case the $\Gamma_n/\,\Gamma_p$ is considerably increased by the
addition of the OKE mechanism and becomes larger than 0.6 (see Table \ref{tab:riskaon}).


Similar analyses can also be repeated for the corresponding model configurations in
which either final SRC or both initial and final SRC contributions are neglected. The shapes of the spectra are quite
insensitive to the presence or absence of short range correlations, as implemented in our model. On
the other hand, the global size of the spectra and the corresponding integrated
quantities $\Gamma_{p,n}$ are influenced, sometimes significantly, by 
SRC effects. 

\subsection{Angular spectra}\label{sec:angspec}

\begin{figure}[htbp]
\begin{center}
\includegraphics[width=12cm]{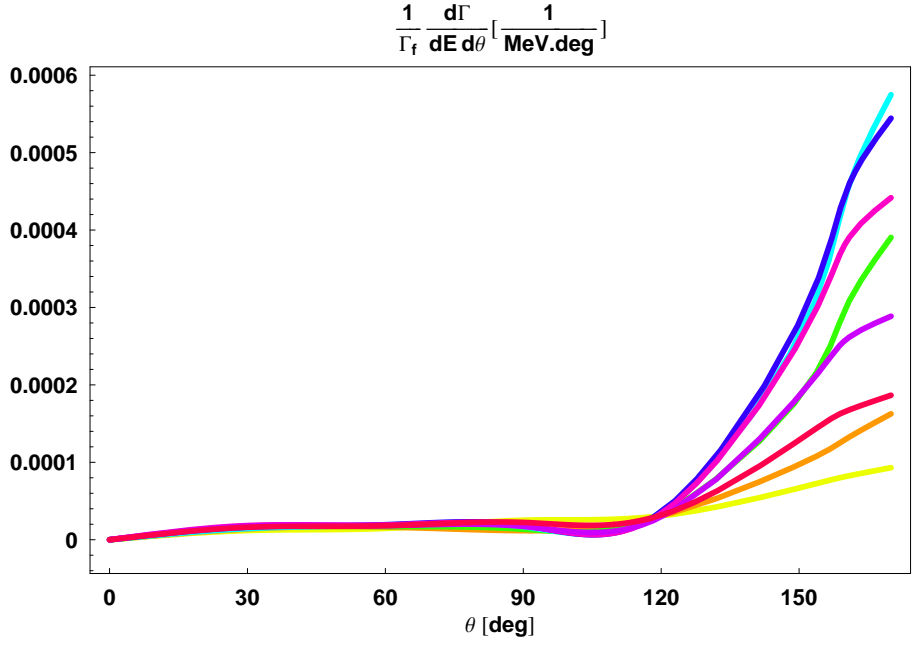}
\caption{\small{Double-differential spectra (normalized to
$\Gamma_\Lambda^{free}$) for the $\Lambda p\to np$ decay channel as a function
of the relative angle $\theta$ between the momenta of the two
outgoing nucleons for different values of the kinetic energy of the outgoing
proton $T_p \equiv E$. Line convention: $E=20$ MeV (orange), $E=36$ MeV (green),
$E=54$ MeV (cyan), $E=68$ MeV (blue), $E=84$ MeV (magenta), $E=100$ MeV (purple),
$E=116$ MeV (red), $E=128$ MeV (yellow).
Calculations are performed with $PV{^\prime}$ couplings and include $\pi+K$
contributions as well as initial and final SRC.}}
\label{fig:AparzP}
\end{center}
\end{figure}

\begin{figure}[htbp]
\begin{center}
\includegraphics[width=12cm]{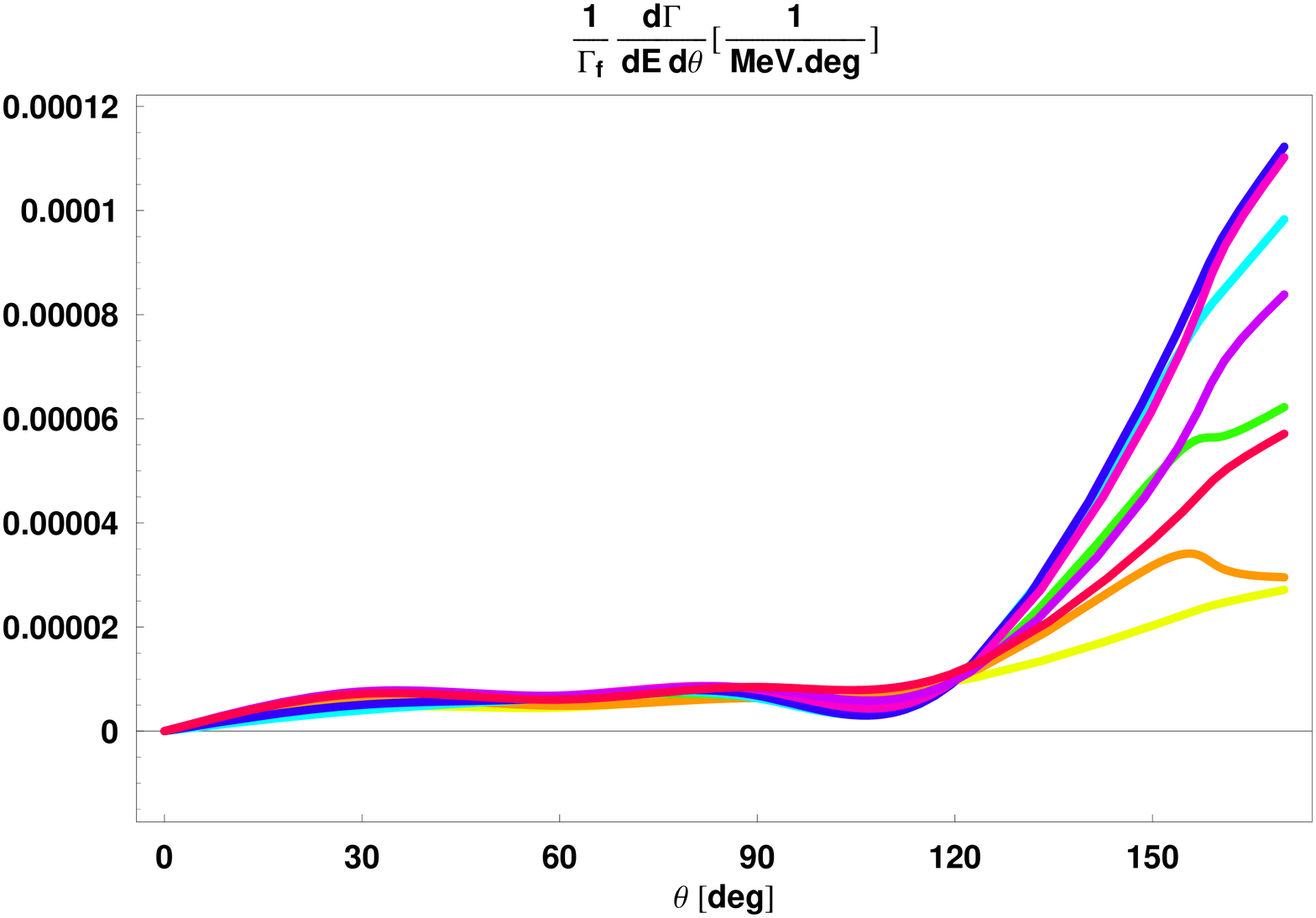}
\caption{\small{The same as in Fig. \ref{fig:AparzP}, but for neutron angular
spectra of the $\Lambda n\to nn$ decay channel.}}
\label{fig:AparzN}
\end{center}
\end{figure}

In this Section we analyze the angular distributions predicted by our model.
In Figs. \ref{fig:AparzP}-\ref{fig:AparzN} the double-differential spectra,
${\rm d}\Gamma_{p,n}/{\rm d}E\,{\rm d}\theta$, for the $\Lambda p\to np$
and $\Lambda n\to nn$  decay channels at different values of the proton  and
neutron kinetic energy, $T_{p,n} \equiv E$, are displayed as a function of the
relative angle $\theta $ between the momenta of the two outgoing nucleons.
The spectra are calculated with $PV{^\prime}$ vertices, including both OPE and
OKE contributions as well as initial and final SRC. The angular spectra for proton- and
neutron-induced channels exhibit a similar behavior. In both figures all the
curves, corresponding to  different values of the kinetic energy of the
outgoing nucleon, are  strongly peaked at high angles, especially for
 $150^{o}\leq\theta\leq180^{o}$. This agrees with the fact that
the elementary process driving the decay is a two-body $\Lambda N\to nN$
interaction, which preferentially yields back-to-back final nucleons.
Distortion effects produced by the nucleon-nucleus optical potential, however,
tend to smear the angular distributions, thus increasing the probability of
emitting nucleons in non back-to-back configurations and at small angles.
In the angular region between $0^{o}$ and $120^{o}$ all the spectra have
similar shapes and comparable sizes, while in the back-to-back region they
suddenly increase and differentiate among each others. The angular spectra
associated with central energies (52, 68, 84 MeV) display a higher peak, while
the curves obtained for energies close to the energy endpoints
(20 and 128 MeV) are much less peaked and definitely smaller in size.


The evidence that the most back-to-back peaked angular spectra are those
pertaining to central energies can be understood if we observe that a proton
or a neutron kinetic energy close to the middle of the available energy range
means that the two final nucleons have approximately the same energy.
Thus, due to the energy-momentum global conservation in the two-body process, the
two outgoing particles are preferentially emitted along opposite directions.
On the other hand, in those cases in which one of the two nucleons carries
away a large part of the available energy and the other one carries the
remaining small amount,  the angular correlation is weaker and the
corresponding angular spectra are flatter.
The angular distributions are smeared by FSI effects. The complex energy
dependent optical potential distorts the wave functions of the outgoing
nucleons and an imbalance between the energies of the two final nucleons also
favours a weakening of their angular correlation and a correspondingly stronger
smearing of the relative angle distribution.

\begin{figure}[htbp]
\begin{center}
\includegraphics[width=12cm]{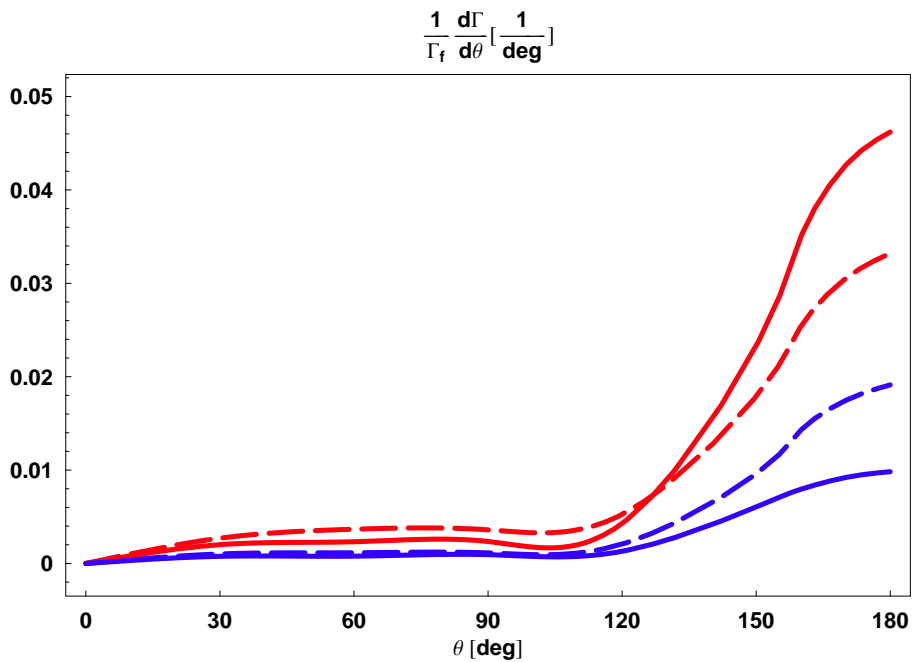}
\caption{\small{Angular spectra (normalized to $\Gamma_\Lambda^{free}$)
integrated over the nucleon kinetic energy $E\equiv T_N$. Calculations are
perfomed with $PV{^\prime}$  couplings and include initial and final SRC. Red (blue)
lines refer to proton (neutron) spectra, as obtained in proton- (neutron-) induced one-body decay
channels. Dashed lines correspond to results obtained by considering only OPE,
solid ones are obtained including also OKE contributions.}}
\label{fig:AtotPV}
\end{center}
\end{figure}

The angular spectra integrated over the kinetic
energy of the outgoing proton (for the $\Lambda p\to np$ channel) and neutron
(for the $\Lambda n\to nn$ channel),  ${\rm d}\Gamma_{p,n}/{\rm d}{\theta}$,
are shown in Fig. \ref{fig:AtotPV}. The results obtained with OPE and $\pi+K$ exchange
are compared in the figure. Calculations have been performed adopting the
$PV{^\prime}$ choice for the vertices and include both initial and final SRC.
By considering only the OPE mechanism we see that the proton
and neutron angular distributions are quite similar in shape, especially in the
back-to-back region, namely from $\theta = 120^{o}-130^{o}$ till $\theta = 180^{o}$, where they present similar peaks.
In the region where $\theta \leq 120^{o}$ the angular spectra are instead approximately flat and strongly reduced in size.
The global size of the proton spectrum is about twice the neutron spectrum one, coherently with the obtained results for the ratio of the
integrated decay rates, $\Gamma_n/\,\Gamma_p$.
The inclusion of the OKE contribution acts in opposite ways on the proton and neutron spectra. The curves are practically unchanged in the region $\theta\leq 120^{o}$, apart from a slight reduction of the proton spectrum. By contrast, in the back-to-back region, the neutron
spectrum is considerably lowered and flattened while the proton spectrum is
correspondingly increased, becoming much more peaked towards higher relative angles.
This behaviour produces the reduction of the $\Gamma_n/\Gamma_p$
ratio shown in Table \ref{tab:riskaon}, when the OKE contribution is included.

\begin{figure}[htbp]
\begin{center}
\includegraphics[width=12cm]{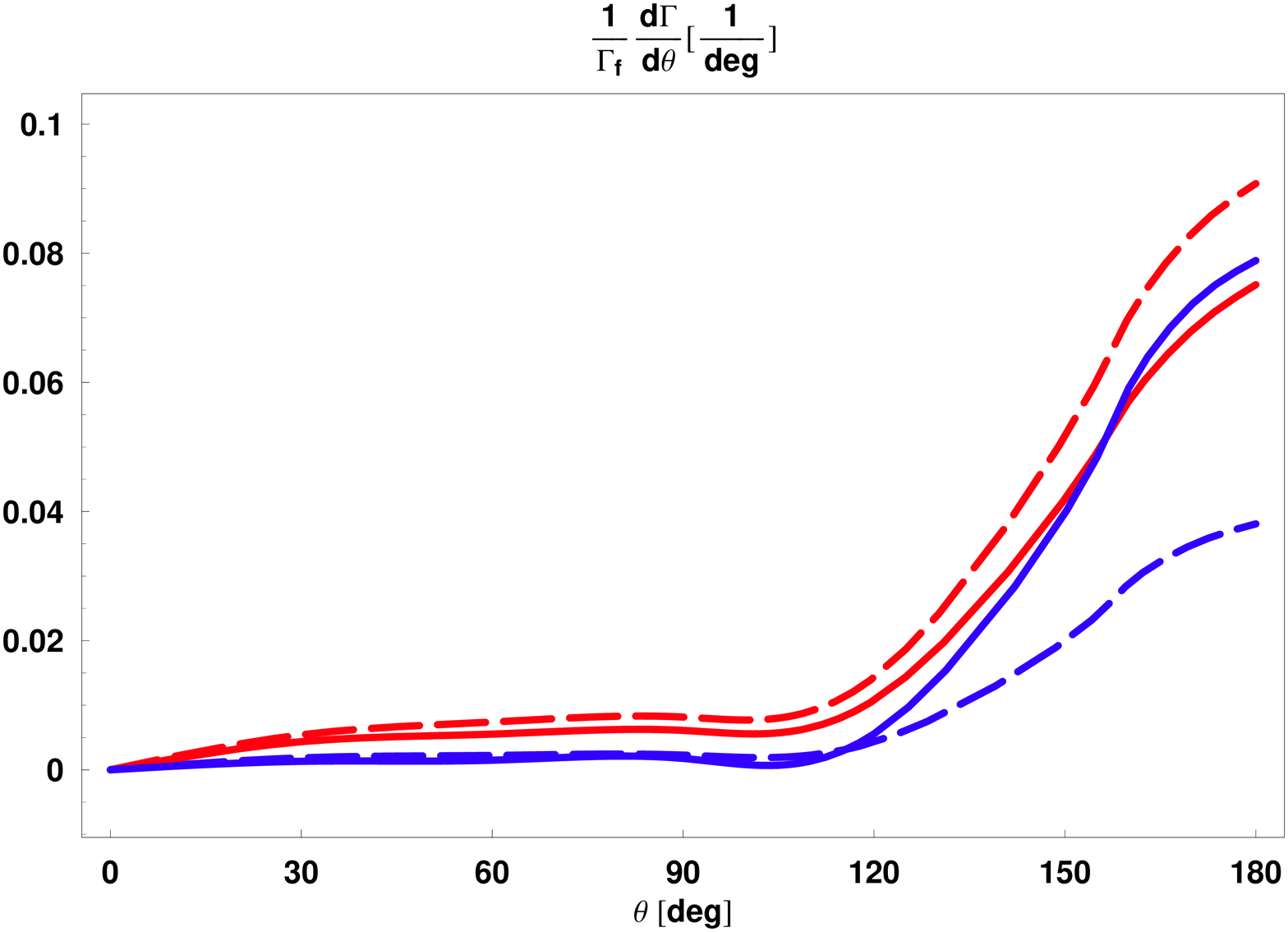}
\caption{\small{The same in Fig. \ref{fig:AtotPV}, but for $PS$ couplings.}}
\label{fig:AtotPS}
\end{center}
\end{figure}

The angular spectra integrated over the kinetic energy of the outgoing proton and neutron calculated with $PS$ couplings (and including initial and final SRC) are  shown in Fig. \ref{fig:AtotPS}. Also in this case the OPE spectra have similar shapes, with the neutron curve
significantly smaller than the proton one in the back-to-back region. On the contrary, the inclusion of the
OKE mechanism induces opposite effects on the angular spectra, when considering $PS$ and $PV{^\prime}$ vertices:
in the $PS$ case the proton distribution is reduced and the neutron one is strongly increased. Thus, the $\Gamma_n/\,\Gamma_p$ ratio is highly enhanced and becomes greater than 0.6 (see Table \ref{tab:riskaon}).


A similar analysis can be repeated for the model configurations in which SRC are completely neglected. The shapes and the relative sizes of the distributions are strictly analogous to the corresponding results obtained when SRC are included.
The conclusion is that, within our model, SRC do not affect in a significant way the shapes and sizes of kinetic energy and angular
spectra. This is coherent with the use of a phenomenological SRC multiplicative, local and energy-independent function.


\section{Summary and conclusions}
\label{conc}

We have presented a relativistic model for the non-mesonic weak decay of the $^{12}_\Lambda C$ hypernucleus.
Over the last years many groups have been deeply involved in studies of hypernuclear physics and developed
different nonrelativistic models that can satisfactorily reproduce all the experimental results. Anyway, the inclusion of many theoretical ingredients, like the full one-meson-exchange potential and many other ones, seems mandatory.
In this paper we have proposed a first attempt to explain, at least in a qualitative way, all the features of the hypernuclear non-mesonic weak decay with a fully relativistic treatment of the weak dynamics, based on the evaluation of Feynman diagrams within a covariant formalism.

We have considered the pseudo-scalar and pseudo-vector prescriptions for the
weak-strong vertices involved in the $\Lambda N\to nN$ elementary process. When considering the standard form of the pseudo-vector vertex,
these two different couplings produce the same nonrelativistic limit, which is anyway inconsistent with the standard nonrelativistic approaches; only the modified pseudo-vector vertex, with a 4-derivative $\partial_{\mu}$ operating on the meson
propagator, allows to obtain, in the nonrelativistic limit, the usual one-pion-exchange potential commonly employed in nonrelativistic calculations. In view of such considerations, we tested our model employing both pseudo-scalar and modified derivative pseudo-vector couplings. In our model short-range correlations are taken into account in a phenomenological way,
through a multiplicative local and energy-independent function. The involved wave functions are 4-spinors obtained
within the framework of Dirac phenomenology in presence of scalar and vector potentials.
Final-state interactions are included in the model by accounting for the
interaction of each one of the two outgoing nucleons with the residual nucleus,
that is implemented by means of a complex relativistic nucleon-nucleus optical potential.
The main effect of the optical potentials is to produce a damping  of the
kinetic energy and angular spectra and to smear the two-body reaction kinematical correlations.

Great care has been devoted to the decay dynamics. The pseudo-vector couplings yield
predictions in reasonable agreement with the non-mesonic weak decay experimental rates, whereas  visible discrepancies
appear when the pseudo-scalar coupling is chosen.
The role of one-pion-exchange and one-kaon-exchange diagrams has been carefully investigated.
When using pseudo-vector couplings, the inclusion of the $K$-exchange is helpful in view of a comparison with
the experimental determinations of the $\Gamma_n/\Gamma_p$ ratio.
On the contrary, if we adopt the pseudo-scalar vertices, the $K$-exchange gives puzzling big values.

The role of initial short-range correlations is only moderate and especially visible in the
total decay rates, which are reduced by about 20-25\%, whereas $\Gamma_n/\Gamma_p$ is
much less sensitive to such a theoretical ingredient. The additional 
consideration of final SRC does not seem to introduce significant modifications, 
when initial SRC are already taken into account. 
We acknowledge that the particular implementation of SRC here adopted, namely 
in terms of local multiplicative functions inspired by non-relativistic 
calculations, could be unsuitable in the context of a fully relativistic 
calculation. Our choice to include such ingredients has been motivated by the 
relevance usually attributed to these correlations in non-relativistic 
hypernuclear decay calculations; the great care here devoted to the problem of 
selecting the right covariant vertices structure in combination with these 
non-relativistic SRC functions was actually aimed at minimizing the impact of 
the theoretical uncertainty introduced by such phenomenological inputs. 
Clearly, modeling short range baryon-baryon correlations directly in the 
relativistic framework, e.g. by considering box diagrams involving heavy mesons 
(typically the $\omega$), would definitely be a better strategy, though much 
more demanding from a calculational viewpoint: we believe this topic deserves 
deeper investigation and our calculation can be a starting point in view of 
similar generalizations.

In contrast with the predictions of nonrelativistic calculations, our model produces significantly high values of the $\Gamma_n/\Gamma_p$ ratio, especially when we only include the one-pion-exchange contribution. As a consequence, our results for $\Gamma_n/\Gamma_p$, both for $\pi$-and $(\pi + K)$-exchange contributions, are close to the experimental measurements,
without any apparent need to include other pseudo-scalar and vector mesons,  that are usually accounted for by the full OME models,
or to resort to more refined FSI models.

It is not easy to understand why the obtained results are so different from well-established non-relativistic predictions.
Due to the high energies involved and short distances probed, the role of relativity could really be important in hypernuclear non-mesonic weak decay; unfortunately, no work to date could demonstrate this 
point with certainty.
A direct comparison between our model and standard nonrelativistic calculations is beyond the scope of the present investigation, also because it would be very difficult to establish the basis for a direct and unambiguous comparison. A nonrelativistic reduction would imply first of all to drop the lower components of the Dirac spinors and apply the proper normalization.
In addition, the relativistic energy $E$ should be put equal to $M$.
However, this is by no means a nonrelativistic reduction, as relativity is
directly included in the vertices and the propagators of the Feynman diagrams,
and the nuclear current operators still involve the Dirac
scalar and vector potentials, which can produce large differences between the results.

We are aware that, at the present stage of development of our work, we cannot derive
any definite conclusion about the relevance or the usefulness of a fully relativistic formalism
to describe the short-range strong-weak dynamical mechanism driving hypernuclear
non-mesonic decay. We are considering the opportunity to better describe
the final state of the decay process, by refining the treatment of short-range strong
correlations between the two outgoing nucleons, here simply described by means of a phenomenological non-relativistic function, and thus evaluating a globally correlated relativistic wave function for the final nucleon pair.
Another possible improvement is related to the dynamics of the model, and requires the implementation of the exchange of vector mesons, thus exploring the effects of these additional contributions on the integrated observables and on the simulated spectra.
Anyway, we think that the results of our model represent an additional source of information and a partly new theoretical perspective, which may deserve attention and further investigation.


\begin{acknowledgments}

We are grateful to W.M. Alberico and G. Garbarino for useful discussions and their valuable advice.

\end{acknowledgments}



\end{document}